\documentclass[a4paper,fleqn]{cas-dc}

\usepackage[authoryear]{natbib}
\usepackage{amssymb}
\usepackage{amsmath}
\usepackage{tikz}
\usepackage{caption}
\usepackage{subcaption}

\newcommand{\incr}{\,\mathrm{d}}
\newcommand{\set}[1]{\{#1\}}
\newcommand{\diff}[2]{\frac{\mathrm{d}{#1}}{\mathrm{d}{#2}}}

\newcommand{\ndiff}[3][]{\frac{\mathrm{d}^{#1}{#2}}{\mathrm{d}{#3}^{#1}}}

\newcommand{\R}{\mathbb R}

\newtheorem{remark}{Remark}

\usepackage{tikz}
\usetikzlibrary{positioning}
\usetikzlibrary{calc}
\usetikzlibrary{shapes.geometric}
\usetikzlibrary{angles}
\usetikzlibrary{decorations.pathreplacing}
\usetikzlibrary{calligraphy}

\newcommand{\valveheight}{ 7pt}
\newcommand{\valvewidth} {10pt}

\tikzstyle{box} = [
    draw,                       
    minimum width   =  25pt,    
    minimum height  =  25pt,    
    rounded corners =   5pt]    
    
\tikzstyle{bigbox} = [
    draw,                       
    minimum width   =  55pt,    
    minimum height  =  55pt,    
    rounded corners =   5pt]    
    
\tikzstyle{longbox} = [
    draw,                       
    minimum width   = 125pt,    
    minimum height  =  25pt,    
    rounded corners =   5pt]    

\begin{document}
    \let\WriteBookmarks\relax
    \def\floatpagepagefraction{1}
    \def\textpagefraction{.001}
    
\shorttitle{Mathematical meal models}

\shortauthors{JB Jørgensen et~al.}

\title[mode = title]{Mathematical meal models for simulation of human metabolism}




\author[DTUComp]{Tobias K. S. Ritschel}[orcid=0000-0002-5843-240X]
\author[DTUComp]{Asbjørn Thode Reenberg}[orcid=0000-0003-0015-7107]
\author[DTUComp]{Peter Emil Carstensen}
\author[DTUComp]{Jacob Bendsen}

%
\author[DTUComp]{John Bagterp Jørgensen}[orcid=0000-0001-7511-2910]

\cormark[1]


\ead{jbjo@dtu.dk}



\affiliation[DTUComp]{organization={Department of Applied Mathematics and Computer Science, Technical University of Denmark},
    addressline={\\Matematiktorvet, Building 303B}, 
    city={Kgs. Lyngby},
    postcode={DK-2800}, 
    country={Denmark}}

\cortext[DTUComp]{Corresponding author}



    \begin{abstract}
    We present and critically discuss five commonly used mathematical models of the meal glucose rate of appearance in humans. Such models are key to simulation of the metabolism in healthy people, people with diabetes, and obese people, and they are central to developing effective treatments and prevention strategies. Furthermore, we discuss important aspects of systematic mathematical modeling of human metabolism, including meal consumption modeling, stoichiometry and reaction kinetics, and general-purpose model components.
\end{abstract} 
    
    \begin{highlights}
    \item Critical review and comparison of existing meal models for dynamic simulation
    \item Discussion of systematic mathematical modeling of gastrointestinal glucose absorption
    \item General-purpose model components for mathematical modeling of human metabolism
\end{highlights} 
    
\begin{keywords}
Gastrointestinal glucose absorption
\sep Mathematical modeling
\sep Simulation
\sep Diabetes
\sep Obesity
\end{keywords} 
    
    \maketitle
    
    \section{Introduction}
Mathematical modeling and simulation of the human metabolism are central to treating and preventing two of the major pandemics of the 21\textsuperscript{st} century; diabetes and obesity~\citep{Pattaranit:vandenBerg:2008}. Model-based simulation can both support scientific developments within physiology, help to improve drug development processes~\citep{Huang:etal:2009}, and be used directly in support tools, e.g., in automated insulin delivery systems for people with diabetes~\citep{Lal:etal:2019}.
Specifically, modeling is a key component of virtual clinical trials~\citep{Reenberg:etal:2022b, Ritschel:etal:2022}, rigorous mathematical analysis~\citep{Cohen:Li:2021}, and of model-based algorithms for, e.g., monitoring, prediction, control, and optimization~\citep{Boiroux:etal:2018b}.

The human metabolism is a complex set of chemical reactions that are responsible for sustaining life. Their purposes are to 1)~digest food, 2)~convert the energy in the food into a form that can be used in cellular processes, 3)~convert food into building blocks for nucleic acids, proteins, carbohydrates, and lipids, 4)~transport substances into and between cells, and 5)~eliminate waste from metabolic processes.
Food digestion and absorption are particularly important to mathematical modeling in diabetes and obesity~\citep{Gouseti:etal:2019, LeFeunteun:etal:2020}.
Many models describe the dynamics of glucose and insulin and disregard other macronutrients (fat and protein) and hormones (e.g., glucagon, ghrelin, and incretins). Furthermore, it is common to describe meal glucose absorption using simple algebraic relations~\citep{Silber:etal:2010} or to only consider intravenous glucose injection~\citep{Silber:etal:2007}. However, models that include the dynamics of glucagon~\citep{Adams:Lasseigne:2018}, ghrelin~\citep{Barnabei:etal:2022}, and incretins~\citep{Jauslin:etal:2007}, as well as the absorption of other macronutrients~\citep{Sicard:etal:2018}, have also been proposed. Recently, \citet{Pompa:etal:2021} compared three models commonly used in diabetes using a simulation study. However, they conclude that it is not possible to determine which of the models that is more physiologically accurate based on simulations alone. \citet{Noguchi:etal:2014} propose a model which accounts for the digestion and absorption of carbohydrates based on the glycemic index and carbohydrate bioavailability. \citet{Moxon:etal:2016} propose three models which include transport along the small intestine. Later, both \citet{Noguchi:etal:2016} and \citet{Moxon:etal:2017} extended their respective models with an upper bound on the glucose rate of appearance in the blood stream.
We refer to the reviews by~\citet{Smith:etal:2009}, \citet{Palumbo:etal:2013}, and~\citet{Huard:Kirkham:2022} for further information on models in the literature.

Apart from the physiological phenomena included in the models, there are several differences between the underlying mathematical formulations. Some models are purely compartmental~\citep{DeGaetano:etal:2013} and described only by ordinary differential equations~(ODEs). Others also use partial differential equations~(PDEs), e.g., to describe the transport through the small intestine~\citep{Moxon:etal:2016}.
Similarly, there are models that represent delays exactly~\citep{Contreras:etal:2020, Cohen:Li:2021} using delay differential equations~(DDEs), and, in other cases, they are approximated~\citep{Alskar:etal:2016}.
%
Furthermore, meal consumption can either be represented as a finite flow rate of nutrients~\citep{Hovorka:etal:2004} or as instantaneous~\citep{DallaMan:etal:2014}. Finally, while most models are deterministic, some also include stochasticity (uncertainty), e.g., to model variations in the meal size and consumption time~\citep{Chudtong:DeGaetano:2021}. These different mathematical formulations are also discussed in the review by~\citet{Makroglou:etal:2006}.

In this work, we present a critical discussion of five commonly used mathematical models of meal glucose absorption: 1)~the model proposed by~\citet{Hovorka:etal:2004}, 2)~the UVA/Padova model presented by~\citet{DallaMan:etal:2006, DallaMan:etal:2007}, 3)~the SIMO model described by~\citet{DeGaetano:etal:2013} and used in the revised Sorensen model by~\citet{Panunzi:etal:2020}, 4)~the model by~\citet{Alskar:etal:2016}, and 5)~a model which represents the stomach as a continuous stirred-tank reactor (CSTR) and the small intestine as a plug-flow reactor (PFR)~\citep{Moxon:etal:2016, Moxon:etal:2017}. In the last model, we compare different models of the opening and closing of the pylorus valve, which connects the stomach to the duodenum in the small intestine. Furthermore, we discuss general aspects of mathematical modeling relevant to the human metabolism (representation of meals, delays, and general modeling components).

The remainder of the paper is structured as follows. In Section~\ref{sec:simulation}, we discuss the simulation of mathematical meal models, and in Section~\ref{sec:models:components}, we present modeling components that are relevant to modeling of the human metabolism in general. In Section~\ref{sec:meal:models}, we present the five models of meal glucose absorption mentioned above, and in Section~\ref{sec:discussion}, we discuss and compare them based on simulations. Finally, we present conclusions in Section~\ref{sec:conclusions}. 
    \section{Mathematical models and simulation}\label{sec:simulation}
We consider mathematical models of the meal nutrient absorption in the gastrointestinal tract, which are in the form of initial value problems involving ODEs:
%
%
\begin{align}\label{eq:IVP:ODE}
    \dot x(t) &= f(x(t), d(t), p_f), & x(t_0) &= x_0.
\end{align}
Here, $t$ is time, $x$ are the states, $d$ are the meal inputs, and $p_f$ are the parameters in the model, $f$. $x_0$ are the states at time $t_0$. The states constitute the minimal amount of information necessary for simulating the future evolution of the system~\eqref{eq:IVP:ODE}.

The outputs, $y$, are described by the function
\begin{equation}
\label{eq:NonlinearOutputFunction}
    y(t) = g(x(t), p_g),
\end{equation}
where $p_g$ is a parameter vector. The purpose of the model is to describe the relation between the outputs, $y$, and 1)~the meal inputs, $d$, and 2)~the parameters, $p_f$ and $p_g$, which are specific to each person (and possibly also to each meal).

\begin{remark}
Models that contain PDEs or DDEs can be approximated by models that only contain ODEs by means of spatial discretizations and delay approximations, respectively.
\end{remark}

\subsection{Typical structure of nonlinear models}
Nonlinear models of meal nutrient absorption are often more structured than the system~\eqref{eq:IVP:ODE}. Specifically, many nonlinear models are affine in the meal inputs:
\begin{equation}
\begin{split}
    \dot x(t) &= f(x(t), d(t), p_f) \\
    &= f_x(x(t), p_{f_x}) + f_d(x(t), p_{f_d}) d(t), \quad x(t_0) = x_0.
\label{eq:IVP:ODE:NonlinearAffineInputModel}
\end{split}
\end{equation}
The first term describes the internal dynamics of the meal absorption and the second term describes the direct effect of the meal inputs on the states, e.g., the relation between the amount of glucose in the meal and in the stomach. The parameter vectors $p_{f_x}$ and $p_{f_d}$ contain the same parameters as $p_f$.

\subsection{Linear models}
Several meal models are linear in the states, $x$, and the meal inputs, $d$:
\begin{subequations}
\label{eq:LinearStateSpaceModel}
\begin{align}
    \dot x(t) &= A_c(p_{f_x}) x(t) + B_c(p_{f_d}) d(t), \quad x(t_0) = x_0, \label{eq:LinearStateSpaceModel:Dynamics} \\
    y(t) &=  C_c(p_g) x(t).
\end{align}
\end{subequations}
The subscript $c$ on the system matrices, $A_c$, $B_c$, and $C_c$, indicate that it is a \emph{continuous-time} linear state space model (as opposed to a \emph{discrete-time} state space model).

\begin{remark}
The linear state space model~\eqref{eq:LinearStateSpaceModel} is a special case of the nonlinear model \eqref{eq:IVP:ODE}--\eqref{eq:NonlinearOutputFunction} where
\begin{subequations}
\begin{align}
    f(x(t), d(t), p_f) &= A_c(p_{f_x}) x(t) + B_c(p_{f_d}) d(t), \\
    g(x(t), p_g) &= C_c(p_g) x(t).
\end{align}
\end{subequations}
The dynamical equation in the linear state space model~\eqref{eq:LinearStateSpaceModel:Dynamics} is also a special case of the meal input-affine model~\eqref{eq:IVP:ODE:NonlinearAffineInputModel} where
\begin{subequations}
\begin{align}
    f_x(x(t), p_{f_x}) &= A_c(p_{f_x}) x(t), \\
    f_d(x(t), p_{f_d}) &= B_c(p_{f_d}).
\end{align}
\end{subequations}
\end{remark}

\subsection{Meal inputs}\label{sec:simulation:meal:inputs}
Some models of meal nutrient absorption represent the meal inputs as flow rates, i.e., as step functions, and others represent them as instantaneous, i.e., as impulses.

\subsubsection{Step inputs}
When the meal inputs are represented using step functions, they are described by
%
\begin{equation}
    d(t) = d_k = D_k/\Delta t, \quad t_k \leq t < t_{k+1},
\end{equation}
where $D_k$ is the total meal size ingested in the interval $[t_k, \,t_{k+1}[$ and $\Delta t = t_{k+1}-t_k$. The response to a sequence of piecewise constant meal inputs of sizes $\set{D_k}_{k=0}^{M-1}$ may be simulated by setting $x(t_0) = x_0$ and solving the $M$ initial value problems
\begin{subequations}
\begin{align}
    x(t_k) &= x_k, \\
    \dot x(t) &= f(x(t), d_k, p_f), & t_{k} &\leq t < t_{k+1}, \\
    x_{k+1} &= x(t_{k+1}),
\end{align}
\end{subequations}
for $k = 0,1, \ldots, M-1$. The result is the sequence of states $\set{x_k}_{k=0}^M$ that may be used to compute the corresponding sequence of outputs $\set{y_k}_{k=0}^M$ from \eqref{eq:NonlinearOutputFunction}.

\subsubsection{Impulse inputs}
For this type of meal input model, we only consider the input-affine model~\eqref{eq:IVP:ODE:NonlinearAffineInputModel}.
A single meal of size $D$, which is consumed instantaneously at time $t_0$, can be represented as an impulse,
\begin{equation}\label{eq:single:meal:impulse}
    d(t) = D \delta(t - t_0),
\end{equation}
using the Dirac delta function, $\delta(t)$. 
We denote by $t_0^-$ the time $t_0$ before the impulse and by $t_0^+$ the time $t_0$ immediately after the impulse. The impulse function has three relevant properties:
\begin{subequations}
\begin{align}
    d(t_0) &= \infty, \\
    d(t) &= 0, & t_0 &< t,  \\
    \int_{t_0^-}^{t_0^+} d(t) \incr t &= D.
\end{align}
\end{subequations}
Consequently, the states immediately before and after the impulse are
\begin{subequations}
\begin{align}
x(t_0^-) &= x_0^- = x_0, \\
x(t_0^+) &= x_0^+ = x_0^- + f_d(x_0^-, p_{f_d}) D.    
\end{align}
\end{subequations}
Therefore, the initial value problem~\eqref{eq:IVP:ODE:NonlinearAffineInputModel} with the meal input function~\eqref{eq:single:meal:impulse} can be simulated by solving the initial value problem
\begin{equation}
    \dot x(t) = f(x(t), 0, p_f) = f_x(x(t), p_{f_x}), \quad x(t_0^+) = x_0^+,
\end{equation}
for $t \geq t_0^+$. The corresponding output, $y(t)$, computed by \eqref{eq:NonlinearOutputFunction} is called the impulse response of the system~\eqref{eq:IVP:ODE:NonlinearAffineInputModel} and~\eqref{eq:NonlinearOutputFunction} to the meal impulse, $D$, provided that $x(t_0) = x_0 = x_{ss}$ is a steady state, i.e., that $x_{ss}$ satisfies $f(x_{ss}, 0, p_f) = f_x(x_{ss}, p_{f_x}) = 0$.

Multiple instantaneous meals (i.e., impulses) of sizes $\set{D_k}_{k=0}^{M-1}$ at times $\set{t_k}_{k=0}^{M-1}$ can be represented by the input function
\begin{equation}
\label{eq:MultipleImpulseInputFunction}
    d(t) = \sum_{k=0}^{M-1} D_k \delta(t - t_k).
\end{equation}
This input function has the three properties
\begin{subequations}
\begin{align}
    d(t_k) &= \infty, \\
    d(t) &= 0, & t_k &< t < t_{k+1},  \\
    \int_{t_{k}^-}^{t_k^+} d(t) \incr t &= D_k.
\end{align}
\end{subequations}
The definitions of $t_k^-$ and $t_k^+$ are analogous to those of $t_0^-$ and $t_0^+$, respectively. Because of these properties, the system~\eqref{eq:IVP:ODE:NonlinearAffineInputModel} with the multiple meal impulse input function \eqref{eq:MultipleImpulseInputFunction} may be simulated by using that $x_0^- = x_0$ and solving the $M$ initial value problems
\begin{subequations}
\begin{align}
    x(t_k^+)  &= x_k^- + f_d(x_k^-,p_{f_d}) D_k, \\
    \dot x(t) &= f_x(x(t),p_{f_x}),  & t_k^+ &< t < t_{k+1}^-, \\
    x_{k+1}^- &= x(t_{k+1}^-),
\end{align}
\end{subequations}
for $k= 0, \ldots, M-1$. As mentioned previously,
\begin{equation}
    f_x(x(t), p_{f_x}) = f(x(t), 0, p_f),
\end{equation}
which means that the simulation can be carried out with the general dynamic model~\eqref{eq:IVP:ODE} using $d(t) = 0$ for $t_k < t < t_{k+1}$. 
    \section{Model components}\label{sec:models:components}
As the human metabolism is a set of chemical reactions, the gastrointestinal tract can be modeled mathematically using modeling techniques from chemical reaction engineering. In this section, we briefly outline a systematic modeling approach using stoichiometry and reaction kinetics in combination with ideal CSTRs and PFRs. Furthermore, delays play an important role in metabolic modeling~\citep{Voit:2017}, and we describe several mathematical models and approximations of delayed signals.


\subsection{Stoichiometry and reaction kinetics}\label{sec:stoichiometry}
Consider a set of molecules $\mathcal{C}$ which are involved in a set of reactions $\mathcal{R}$ in the human metabolism. Let $S \in \R^{n_r \times n_c}$ be the matrix of stoichiometric coefficients for this set of reactions and molecules. $n_c$ is the number of molecules and $n_r$ is the number of reactions. Let $c$ be the vector of concentrations such that we can express the rate vector, $r$, for this set of reactions as the function 
\begin{equation}
    r = r(c).
\end{equation}
Consequently, the production rate vector for the molecules can be expressed as
\begin{equation}
    R = S' r.
\end{equation}
This general way of expressing the production rate, $R$, is useful because it only requires the specification of the chemical reaction stoichiometry (and the corresponding stoichiometric matrix, $S$) as well as the corresponding expression for the reaction rates, $r = r(c)$.
\subsection{CSTR}\label{sec:models:components:cstr}
Any part of the gastrointestinal tract where transport phenomena (i.e., advection and diffusion) are negligible can be represented as a CSTR. The mass balance for a CSTR is
\begin{equation}
    V \dot c = (c_{in} - c) F + R V,
\end{equation}
where $V$ is volume, $c$ is concentration, $c_{in}$ is the inflow concentration, $F$ is the volumetric in- and outflow rate, and $R$ is the production rate. The volume is assumed to be constant, and the model can be reformulated as an ODE:
\begin{equation}
    \dot c = (c_{in} - c) F/V + R.
\end{equation}
\subsection{PFR}\label{sec:models:components:pfr}
The parts of the gastrointestinal tract where advective and diffusive transport phenomena are significant can be described as PFRs. A PFR is cylindrical and the concentration, $c = c(t, z, r, \theta) = c(t, z)$, only changes along the transport direction, $z$, i.e., it is constant along the radial and angular coordinates, $r$ and $\theta$.

The spatiotemporal evolution of the concentration is described by the PDE
\begin{equation}\label{eq:pfr}
    \partial_t c = - \partial_z N + R + Q,
\end{equation}
where $N$ is flux, $R$ is the production rate, and $Q$ is a source term. The flux is the sum of an advection term, $N_a$, and a diffusion term, $N_d$:
\begin{equation}
    N = N_a + N_d.
\end{equation}
These terms are
\begin{subequations}\label{eq:PFR:Flux:Terms}
\begin{align}
    \label{eq:PFR:Flux:Terms:Advection}
    N_a &= v c, \\
    \label{eq:PFR:Flux:Terms:Diffusion}
    N_d &= -D_c \partial_z c,
\end{align}
\end{subequations}
where $v$ is velocity and $D_c$ is the diffusion coefficient. The expression~\eqref{eq:PFR:Flux:Terms:Diffusion} is called Fick's law.



\subsection{Delays}
Here, we describe different formulations and approximations of a model, where $y$ is equal to the input signal~$u$ delayed by $\tau_d$, i.e.,
\begin{align}
\label{eq:TimeDelayModelYU}
    y(t) &= u(t - \tau_d).
\end{align}
The Laplace transform of \eqref{eq:TimeDelayModelYU} is
\begin{subequations}\label{eq:Laplace:TimeDelayModelYU}
\begin{align}
    \label{eq:Laplace:TimeDelayModelYU:YGU}
    Y(s) &= G(s) U(s), \\
    \label{eq:Laplace:TimeDelayModelYU:G}
    G(s) &= e^{-\tau_d s}.
\end{align}
\end{subequations}

Alternatively, the system~\eqref{eq:TimeDelayModelYU} can be formulated as a series of $M$ systems with smaller time delays:
\begin{align}
    y_i(t) &= y_{i-1}(t - \tau_d/M), & i &= 1, \ldots, M,
\end{align}
where
\begin{subequations}
\begin{align}
    y_0(t) &= u(t), \\
    y(t) &= y_M(t).
\end{align}
\end{subequations}
The Laplace transform of this series of systems are
\begin{subequations}
\label{eq:Laplace:TimeDelayModelYU:Multiple}
\begin{align}
    Y_i(s) &= G_i(s) Y_{i-1}(s), & i &= 1, \ldots, M, \\
    G_i(s) &= e^{-(\tau_d/M) s},
\end{align}
\end{subequations}
where
\begin{subequations}
\begin{align}
    Y_0(s) &= U(s), \\
    Y(s)   &= Y_M(s).
\end{align}
\end{subequations}
Approximating $G_i$ in~\eqref{eq:Laplace:TimeDelayModelYU:Multiple} will typically result in a lower error than approximating $G$ in~\eqref{eq:Laplace:TimeDelayModelYU} because the delay is smaller. However, the increased accuracy comes at the expense of higher computational cost.
Below, we show different approximations based on a dynamical system in the form
\begin{subequations}
\label{eq:linear:state:space:model}
\begin{align}
    \dot   x(t) &= A_c x(t) + B_c u(t), \\
    \tilde y(t) &= C_c x(t) + D_c u(t).
\end{align}
\end{subequations}

\subsubsection{Lag approximation}
The transfer function in~\eqref{eq:Laplace:TimeDelayModelYU:G} can be approximated by the transfer function of a lag process, i.e.,
\begin{align}
    G(s) &\approx \frac{1}{\tau_d s + 1} = \frac{1/\tau_d}{s + 1/\tau_d} = \frac{P(s)}{Q(s)} = \tilde G(s).
\end{align}
The system matrices in the corresponding linear state space realization, in observable canonical form~\citep[Chap.~3.9]{Hendricks:etal:2008}, are
\begin{subequations}
\begin{align}
    A_c &= - 1/\tau_d, & B_c &= 1/\tau_d, \\
    C_c &= 1, & D_c &= 0.
\end{align}
\end{subequations}

We apply the same approximation to the system~\eqref{eq:Laplace:TimeDelayModelYU:Multiple}:
\begin{align}
    G_i(s)
    &\approx \frac{1}{(\tau_d/M) s + 1} = \frac{M/\tau_d}{s + M/\tau_d} = \frac{P_i(s)}{Q_i(s)} \nonumber \\
    &= \tilde G_i(s).
\end{align}
Again, we consider the corresponding state space realization in observable canonical form. In this case, the system matrices are
\begin{subequations}
\label{eq:lag:multiple:system:matrices}
\begin{align}
    A_{c, ij} &=
    \begin{cases}
        -M/\tau_d, & i = j, \\
         M/\tau_d, & i = j-1, \\
         0, & \text{otherwise},
    \end{cases} \\
    B_{c, i} &=
    \begin{cases}
        M/\tau_d, & i = 1, \\
        0, & \text{otherwise},
    \end{cases} \\
    C_{c, i} &=
    \begin{cases}
        1, & i = M, \\
        0, & \text{otherwise},
    \end{cases} \\
    D_c &= 0,
\end{align}
\end{subequations}
for $i = 1, \ldots, M$ and $j = 1, \ldots, M$.
\subsubsection{Pad{\'{e}} approximation}
The Pad{\'{e}} approximation~\citep{Wei:Hu:Dai:Wang:2016} is another classical way used to approximate time delays. The first-order Pad{\'{e}} approximation of $G(s) = e^{-\tau_d s}$ is
\begin{align}
\label{eq:Laplace:FirstOrderPadeApproximation}
    G(s) &\approx \frac{-(\tau_d/2)s +1}{(\tau_d/2) s + 1} = \frac{- s + 2/\tau_d}{s+2/\tau_d} = \frac{P(s)}{Q(s)} = \tilde{G}(s).
\end{align}
The first-order Pad{\'{e}} approximation~\eqref{eq:Laplace:FirstOrderPadeApproximation} may be used to approximately realize~\eqref{eq:Laplace:TimeDelayModelYU} as the linear state space model~\eqref{eq:linear:state:space:model}, in observable canonical form, with the system matrices
\begin{subequations}
\begin{align}
    A_c &= -2/\tau_d, & B_c &= 4/\tau_d, \\
    C_c &= 1, & D_c &= -1.
\end{align}
\end{subequations}

The Pad{\'{e}} approximation of $G_i$ in~\eqref{eq:Laplace:TimeDelayModelYU:Multiple} is
%
\begin{align}
    G_i(s)
    &\approx \frac{-(\tau_d/(2M))s +1}{(\tau_d/(2M)) s + 1} = \frac{-s + 2M/\tau_d}{s + 2M/\tau_d} \nonumber \\
    &= \frac{P_i(s)}{Q_i(s)} = \tilde{G}_i(s),
\end{align}
and the system matrices in the corresponding state space realization (in observer canonical form) are
\begin{subequations}
\begin{align}
    A_{c, ij} &=
    \begin{cases}
        -2M/\tau_d, & i = j, \\
         (-1)^{i+j+1}4M/\tau_d, & i > j, \\
         0, & \text{otherwise},
    \end{cases} \\
    B_{c, i} &= (-1)^{i+1} 4M/\tau_d, \\
    C_{c, i} &= (-1)^{M+i}, \\
    D_c &= (-1)^M,
\end{align}
\end{subequations}
for $i = 1, \ldots, M$ and $j = 1, \ldots, M$.
\subsubsection{Physical transport delay model}
Delays can also be represented using transport processes. The input signal, $u$, constitutes the boundary condition,
\begin{align}
    c_{in}(t) = u(t),
\end{align}
and the initial boundary value problem
\begin{subequations}
\label{eq:PDE:timedelay}
\begin{align}
    c(t,0) &= c_{in}(t), \\
    \partial_t c &= - v \partial_z c, & t &\geq 0, & 0 &\leq z \leq L,
\end{align}
\end{subequations}
has the analytical solution $y(t) = c(t, L) = c_{in}(t - \tau_d) = u(t - \tau_d)$ with the delay $\tau_d = L/v$.

\begin{remark}
    A left-sided first-order finite difference discretization of the PDE~\eqref{eq:PDE:timedelay}, based on an equidistant grid with $M+1$ nodes, is equivalent to the linear state space model~\eqref{eq:linear:state:space:model} with the system matrices~\eqref{eq:lag:multiple:system:matrices} obtained using a series of $M$ lag approximations.
\end{remark}
\subsubsection{Algebraic delay approximation}\label{sec:models:components:delay:algebraic}
For completeness, we also describe an algebraic delay approximation which is used in the literature, e.g., by~\citet{Alskar:etal:2016}. However, unlike the previous approximations, it is an algebraic expression rather than a linear state space model in the form~\eqref{eq:linear:state:space:model}. Furthermore, it specifically approximates a step in the input function, $u$, whereas the other approximations can be used for arbitrary input functions.

Let $t_s$ denote the time at which the step in $u$ occurs, i.e., $u(t) = 1$ for $t \geq t_s$ and $u(t) = 0$ otherwise. Then, the approximation is
%
\begin{align}
    y(t) &\approx \frac{1}{1 + \exp(-\sigma(t - t_{50}))} = \tilde y(t),
\end{align}
where $t_{50} = t_s + \tau_d$ is the time at which $\tilde y$ is halfway between the value of $u$ before and after the step.

\begin{figure}
    \centering
    \includegraphics[width=\linewidth]{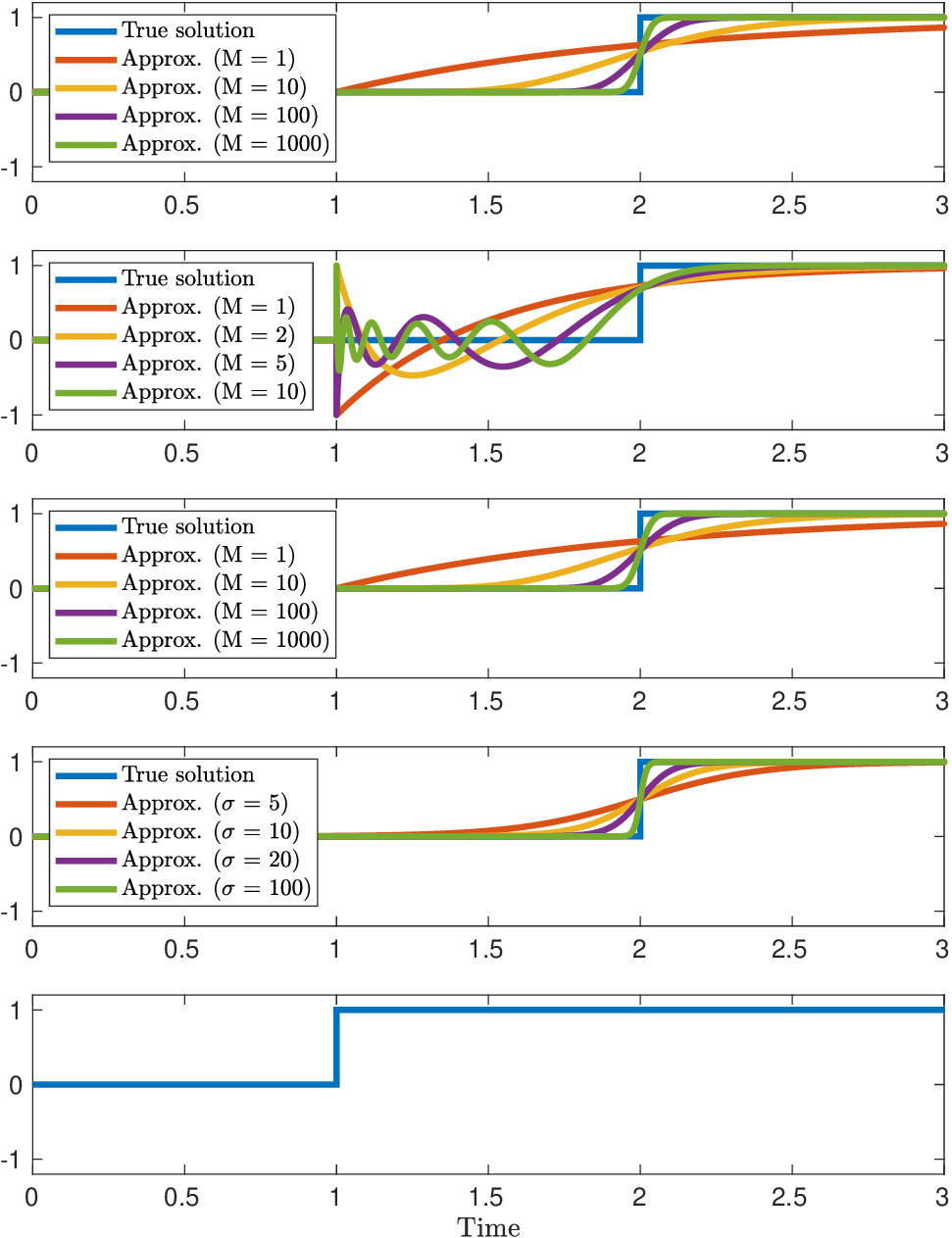}
    \caption{True, $y$, and approximate, $\tilde y$, delays of the input, $u$. Top: Lag approximation. Second from the top: Pad{\'{e}} approximation. Third from the top: Finite difference discretization of physical transport delay model. Fourth from the top: Algebraic delay approximation. Bottom: Input, $u$.}
    \label{fig:delayModels}
\end{figure} 
    \section{Meal models}\label{sec:meal:models}
In this section, we present five commonly used models of glucose absorption in the gastrointestinal tract: The model by~\citet{Hovorka:etal:2004}, the model by~\citet{DallaMan:etal:2006, DallaMan:etal:2007}, the SIMO model by~\citet{DeGaetano:etal:2013}, the model by~\citet{Alskar:etal:2016}, and a CSTR-PFR model based on the ones proposed by~\citet{Moxon:etal:2016, Moxon:etal:2017}. For the CSTR-PFR model, we consider different descriptions of the pylorus sphincter (or valve) which connects the stomach to the small intestine.

Several of the models are linear. Specifically, they are in the form
\begin{subequations}
\label{eq:linear:state:space:meal:model}
\begin{align}
    \dot x(t) &= A_c x(t) + B_c d(t), \\
         y(t) &= C_c x(t),
\end{align}
\end{subequations}
where $d$ is the meal input and $y$ is the glucose rate of appearance in the blood plasma.
Furthermore, in Appendix~\ref{sec:LinearGlucoseRateOfAppearance}, we show that, for some of the models, $y$ is a linear function of the total meal carbohydrate content, $D$.
For brevity of notation, we omit the time dependency in the remainder of this section.

\subsection{Hovorka's model}
\begin{figure}
    \centering
    \begin{tikzpicture}
    \node[box] (D1) {$D_1$};
    
    \node[box, right=30pt of D1] (D2) {$D_2$};
    
    \draw[->] ($(D1.west) - (30pt, 0pt)$) -- (D1)                        node [pos=0.5, anchor=south] {$d$};
    \draw[->] (D1)                        -- (D2)                        node [pos=0.5, anchor=south] {$R_{12}$};
    \draw[->] (D2)                        -- ($(D2.east) + (30pt, 0pt)$) node [pos=0.5, anchor=south] {$R_2$};
\end{tikzpicture}
    \caption{Sketch of the meal model presented by~\citet{Hovorka:etal:2004}.}
    \label{fig:hovorka:diagram}
\end{figure}
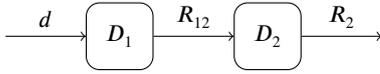

The model by~\citet{Hovorka:etal:2004} contains two compartments, as illustrated in Fig.~\ref{fig:hovorka:diagram}.
The first compartment, $D_1$, describes the amount of glucose in the stomach, and the second compartment, $D_2$, describes the amount of glucose in the small intestine:
\begin{subequations}
\begin{alignat}{3}
    \dot D_1 &= A_G d - R_{12}, \\
    \dot D_2 &= R_{12} - R_2.
\end{alignat}
\end{subequations}
Here, $A_G$ describes the bioavailability of the carbohydrates in the meal, and $R_{12} = R_{12}(D_1)$ is the glucose flow rate between the stomach and the small intestine. Furthermore, $R_2 = R_2(D_2)$ describes the glucose absorption, and the glucose rate of appearance, $R_A = R_A(D_2)$, is a fraction, $f$, of $R_2$:
\begin{subequations}
\begin{align}
    R_{12} &= D_1/\tau_D, \\
    R_2 &= D_2/\tau_D, \\
    R_A &= f R_2.
\end{align}
\end{subequations}
The parameter $\tau_D$ is a time constant, and the model is a linear state space model in the form~\eqref{eq:linear:state:space:meal:model}, where the system matrices are
\begin{align}
    A_c &= \begin{bmatrix} \frac{-1}{\tau_D} & 0 \\ \frac{1}{\tau_D} & \frac{-1}{\tau_D} \end{bmatrix}, &
    B_c &= \begin{bmatrix} A_G \\ 0 \end{bmatrix}, &
    C_c &= \begin{bmatrix} 0 & \frac{f}{\tau_D}\end{bmatrix}.
\end{align}
\subsection{Dalla Man's model}
\begin{figure}
    \centering
    \begin{tikzpicture}
    \node[box] (Qsto1) {$Q_{sto, 1}$};
    
    \node[box, right=30pt of Qsto1] (Qsto2) {$Q_{sto, 2}$};
    
    \node[box, right=30pt of Qsto2] (Qgut) {$Q_{gut}$};
    
    \draw[->] ($(Qsto1.west) - (30pt, 0pt)$) -- (Qsto1)                        node [pos=0.5, anchor=south] {$d$};
    \draw[->] (Qsto1)                        -- (Qsto2)                        node [pos=0.5, anchor=south] {$R_{12}$};
    \draw[->] (Qsto2)                        -- (Qgut)                         node [pos=0.5, anchor=south] {$R_{sto, gut}$};
    \draw[->] (Qgut)                         -- ($(Qgut.east) + (30pt, 0pt)$)  node [pos=0.5, anchor=south] {$R_{gut, pla}$};
\end{tikzpicture}
    \caption{Sketch of the meal model presented by~\citet{DallaMan:etal:2006}.}
    \label{fig:uvapadova:diagram}
\end{figure}
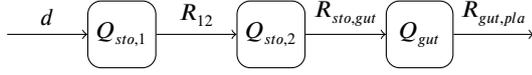

The model by~\citet{DallaMan:etal:2006, DallaMan:etal:2007} is sketched in Fig.~\ref{fig:uvapadova:diagram}, and it contains three compartments: The glucose in the solid and liquid phases of the stomach content, $Q_{sto, 1}$ and $Q_{sto, 2}$, respectively, and the amount of glucose in the small intestine, $Q_{gut}$.
The compartments are described by
\begin{subequations}
\begin{align}
    \dot Q_{sto,1} &= d - R_{12}, \\
    \dot Q_{sto,2} &= R_{12} - R_{sto,gut},  \\
    \dot Q_{gut}   &= R_{sto,gut} - R_{gut, pla},
\end{align}
\end{subequations}
where $R_{12} = R_{12}(Q_{sto, 1})$ is the glucose flow rate between the liquid and solid phase in the stomach, $R_{sto, gut} = R_{sto, gut}(Q_{sto, 1}, Q_{sto, 2}, D)$ is the flow rate between the stomach and the small intestine, and $D$ is the total carbohydrate content of the meal. Furthermore, $R_{gut, pla} = R_{gut, pla}(Q_{gut})$ is the glucose absorption rate, and the glucose rate of appearance in the blood plasma, $R_A = R_A(Q_{gut})$, is a fraction, $f$, of the glucose absorption rate:
\begin{subequations}
\begin{align}
    R_{12} &= k_{gri} Q_{sto, 1}, \\
    R_{sto, gut} &= k_{empt} Q_{sto, 2}, \\
    R_{gut, pla} &= k_{abs} Q_{gut}, \\
    R_A &= f R_{gut, pla}.
\end{align}
\end{subequations}
Here, $k_{gri}$ and $k_{abs}$ are the inverses of time constants, and the gastric emptying rate, $k_{empt} = k_{empt}(Q_{sto}, D)$, is
\begin{subequations}
\begin{align}
    \label{eq:dalla:man:gastric:emptying:rate}
    k_{empt}
    =&\, k_{min} + \frac{k_{max} - k_{min}}{2}\Bigg(\tanh\left(\alpha(Q_{sto} - b D)\right) \nonumber \\
    &- \tanh\left(\beta(Q_{sto} - c D)\right) + 2\Bigg), \\
    \alpha
    =&\, \frac{5}{2 D (1 - b)}, \\
    \beta
    =&\, \frac{5}{2 D c},
\end{align}
\end{subequations}
where $Q_{sto} = Q_{sto}(Q_{sto, 1}, Q_{sto, 2})$ is the total amount of glucose in the stomach:
\begin{equation}
    Q_{sto} = Q_{sto,1} + Q_{sto,2}.
\end{equation}
Furthermore, the parameters $k_{min}$ and $k_{max}$ are the minimum and maximum gastric emptying rates, and $b$ and $c$ are the percentages of $D$ where the magnitude of the derivative of $k_{empt}$ is $\frac{1}{2}(k_{max} - k_{min})$, i.e., at $Q_{sto} = bD$ and $Q_{sto} = cD$. Finally, as $k_{empt}$ in~\eqref{eq:dalla:man:gastric:emptying:rate} is nonlinear in $Q_{sto}$, the model is not in the linear form~\eqref{eq:linear:state:space:meal:model}.

\subsection{The SIMO model}
The SIMO model by~\citet{DeGaetano:etal:2013} contains four compartments, and it is sketched in Fig.~\ref{fig:simo:diagram}. The compartments represent the amounts of glucose in 1)~the stomach, $S$, 2)~the jejunum, $J$, 3)~an artificial delay compartment, $R$, and 4)~the ileum, $L$:
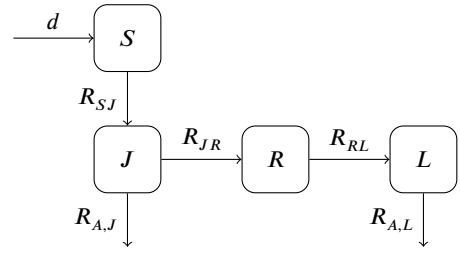
\begin{figure}
    \centering
    \begin{tikzpicture}
    \node[box] (S) {$S$};
    
    \node[box, below=20pt of S] (J) {$J$};
    
    \node[box, right=30pt of J] (R) {$R$};
    
    \node[box, right=30pt of R] (L) {$L$};
    
    \draw[->] ($(S.west) - (30pt, 0pt)$) -- (S)                          node [pos=0.5, anchor=south] {$d$};
    \draw[->] (S)                        -- (J)                          node [pos=0.5, anchor=east ] {$R_{SJ}$};
    \draw[->] (J)                        -- (R)                          node [pos=0.5, anchor=south] {$R_{JR}$};
    \draw[->] (J)                        -- ($(J.south) - (0pt, 20pt)$)  node [pos=0.5, anchor=east ] {$R_{A, J}$};
    \draw[->] (R)                        -- (L)                          node [pos=0.5, anchor=south] {$R_{RL}$};
    \draw[->] (L)                        -- ($(L.south) - (0pt, 20pt)$)  node [pos=0.5, anchor=east ] {$R_{A, L}$};
\end{tikzpicture}
    \caption{Sketch of the SIMO meal model presented by~\citet{DeGaetano:etal:2013}.}
    \label{fig:simo:diagram}
\end{figure}
\begin{subequations}
\begin{align}
    \dot S &= d - R_{SJ}, \\
    \dot J &= R_{SJ} - R_{JR} - R_{A, J}, \\
    \dot R &= R_{JR} - R_{RL}, \\
    \dot L &= R_{RL} - R_{A, P}.
\end{align}
\end{subequations}
Here, the glucose flow rates between the stomach and the jejunum, $R_{SJ} = R_{SJ}(S)$, between the jejunum and the delay compartment, $R_{JR} = R_{JR}(J)$, and between the delay compartment and the ileum, $R_{RL} = R_{RL}(R)$, as well as the glucose absorption rates in the jejunum, $R_{A, J} = R_{A, J}(J)$, and the ileum, $R_{A, L} = R_{A, L}(L)$, are given by
\begin{subequations}
\begin{align}
    R_{SJ}   &= k_{js} S, \\
    R_{JR}   &= k_{rj} J, \\
    R_{RL}   &= k_{lr} R, \\
    R_{A, J} &= k_{gj} J, \\
    R_{A, L} &= k_{gl} L.
\end{align}
\end{subequations}
The coefficients, $k_{js}$, $k_{rj}$, $k_{lr}$, $k_{gj}$, and $k_{gl}$ are the inverses of time constants, and the glucose rate of appearance, $R_A = R_A(J, L)$, is a fraction, $f$, of the total glucose absorption:
\begin{align}
    R_A &= f (R_{A, J} + R_{A, L}).
\end{align}
This model is in the linear form~\eqref{eq:linear:state:space:meal:model}, and the system matrices are given by
\begin{subequations}
\begin{align}
    A_c &=
    \begin{bmatrix}
        -k_{js} &                  0 &       0 &       0 \\
         k_{js} & -(k_{gj} + k_{rj}) &       0 &       0 \\
              0 &            k_{rj}  & -k_{lr} &       0 \\
              0 &                  0 &  k_{lr} & -k_{gl}
    \end{bmatrix}, \\
    B_c &=
    \begin{bmatrix}
        1 \\ 0 \\ 0 \\ 0
    \end{bmatrix}, \\
    C_c &=
    \begin{bmatrix}
        0 & f k_{gj} & 0 & f k_{gl}
    \end{bmatrix}.
\end{align}
\end{subequations}
\subsection{Alsk{\"{a}}r's model}\label{sec:meal:models:alskar}
The model by~\citet{Alskar:etal:2016} contains four compartments representing the amounts of glucose in the stomach, $G_S$, the duodenum, $G_D$, the jejunum, $G_J$, and the ileum, $G_I$, as illustrated in Fig.~\ref{fig:alskar:diagram}. They are described by
\begin{figure}
    \centering
    \begin{tikzpicture}
    \node[box] (GS) {$G_S$};
    
    \node[below=10pt of GS, text width=50pt, align=right] (pylorus) {$R_{SD}$};
    
    \node[box, below=10pt of pylorus] (GD) {$G_D$};
    
    \node[box, right=30pt of GD] (GJ) {$G_J$};
    
    \node[box, right=30pt of GJ] (GI) {$G_I$};
    
    \draw[->] ($(GS.west) - (30pt, 0pt)$) -- (GS)                          node [pos=0.5, anchor=south] {$d$};
    \draw[- ] (GS)                        -- ($(pylorus.center) + (0pt, \valveheight)$);
    \draw[->] ($(pylorus.center) - (0pt, \valveheight)$) -- (GD);
    \draw[->] (GD)                        -- ($(GD.south) - (0pt, 20pt)$)  node [pos=0.5, anchor=east ] {$R_{A, D}$};
    \draw[->] (GD)                        -- (GJ)                          node [pos=0.5, anchor=south] {$R_{DJ}$};
    \draw[->] (GJ)                        -- ($(GJ.south) - (0pt, 20pt)$)  node [pos=0.5, anchor=east ] {$R_{A, J}$};
    \draw[->] (GJ)                        -- (GI)                          node [pos=0.5, anchor=south] {$R_{JI}$};
    \draw[->] (GI)                        -- ($(GI.south) - (0pt, 20pt)$)  node [pos=0.5, anchor=east ] {$R_{A, I}$};
    \draw[->, dashed] (GD) -- ++(-32pt, 0pt) -- ++(0, 30pt) -- ++(13pt, 0pt);
    
    \draw (pylorus.center) -- ++(0.5*\valvewidth, \valveheight) -- ++(-\valvewidth, 0pt) -- (pylorus.center)                                
    -- ++(0.5*\valvewidth, -\valveheight) -- ++(-\valvewidth, 0pt) -- (pylorus.center)                                                      
    -- ++(-\valvewidth, 0pt) -- ++(0pt, \valveheight) to[out=180, in=180, looseness=2] ++(0pt, -2*\valveheight) -- ++(0pt, \valveheight);   
\end{tikzpicture}
    \caption{Sketch of the meal model presented by~\citet{Alskar:etal:2016}.}
    \label{fig:alskar:diagram}
\end{figure}
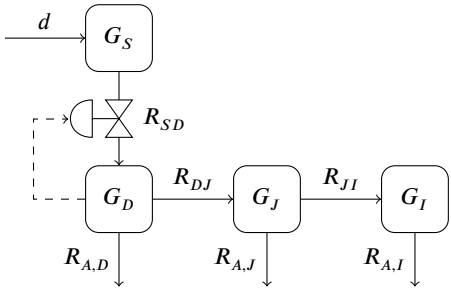
\begin{subequations}
\begin{align}
    \dot G_S &= d - R_{SD}, \\
    \dot G_D &= R_{SD} - R_{DJ} - R_{A, D}, \\
    \dot G_J &= R_{DJ} - R_{JI} - R_{A, J}, \\
    \dot G_I &= R_{JI} - R_{A, I},
\end{align}
\end{subequations}
where the glucose flow rates between the stomach and duodenum, $R_{SD} = R_{SD}(G_S)$, between the duodenum and jejunum, $R_{DJ} = R_{DJ}(G_D)$, and between the jejunum and ileum, $R_{JI} = R_{JI}(G_J)$, are
\begin{subequations}
\begin{align}
    R_{SD} &= k_{SD} \tau G_S, \\
    R_{DJ} &= k_{DJ} G_D, \\
    R_{JI} &= k_{JI} G_J.
\end{align}
\end{subequations}
The inverses of the time constants, $k_{SD} = k_{SD}(G_D)$, $k_{DJ}$, and $k_{JI}$, are
\begin{subequations}\label{eq:alskar:k}
\begin{align}
    \label{eq:alskar:k:sd}
    k_{SD} &= k_w\left(1 - \frac{G_D^\gamma}{IG_{D50}^\gamma + G_D^\gamma}\right), \\
    \label{eq:alskar:k:df}
    k_{DJ} &= \frac{1}{L_D T}, \\
    \label{eq:alskar:k:ji}
    k_{JI} &= \frac{1}{L_J T}.
\end{align}
\end{subequations}
Here, $k_{SD}$ represents the pylorus sphincter, and it is a function of the amount of glucose in the duodenum described using the Hill expression. For $G_D = 0$, $k_{SD}$ is equal to its nominal value, $k_w$, and, as $G_D$ increases, $k_{SD}$ approaches zero. For large values of the Hill coefficient, $\gamma$, $k_{SD}$ has a steep decrease around $IG_{D50}$. Furthermore, $L_D$ and $L_J$ are the relative lengths of the duodenum and jejunum (i.e., fractions of the total length of the small intestine), and $T$ is the transit time through the small intestine. The lag coefficient (used to approximate a time delay) is given by
\begin{align}
    \tau &= \frac{1}{1 + \exp(-\sigma(t - t_{50}))},
\end{align}
as described in Section~\ref{sec:models:components:delay:algebraic}. The parameter $\sigma$ determines the steepness, and $t_{50}$ is the time at which $\tau$ is 0.5. The glucose absorption rates in the duodenum, $R_{A, D} = R_{A, D}(G_D)$, jejunum, $R_{A, J} = R_{A, J}(G_J)$, and ileum, $R_{A, I} = R_{A, I}(G_I)$, are described using Michaelis-Menten expressions, i.e.,
\begin{subequations}
\begin{align}
    R_{A, D} &= \frac{R_{D, \max} G_D}{K_{mG} + G_D}, \\
    R_{A, J} &= \frac{R_{J, \max} G_J}{K_{mG} + G_J}, \\
    R_{A, I} &= \frac{R_{I, \max} G_I}{K_{mG} + G_I},
\end{align}
\end{subequations}
where $K_{mG}$ is the Michaelis constant and $R_{D, \max}$, $R_{J, \max}$, and $R_{I, \max}$ are the maximum glucose absorption rates in the duodenum, jejunum, and ileum, respectively. Finally, the glucose rate of appearance in the blood plasma, $R_A = R_A(G_D, G_J, G_I)$, is a fraction, $F_P$, of the total glucose absorption:
\begin{align}
    R_A &= F_P (R_{A, D} + R_{A, J} + R_{A, I}).
\end{align}
\subsection{CSTR-PFR model}\label{sec:cstr:pfr:meal:model}
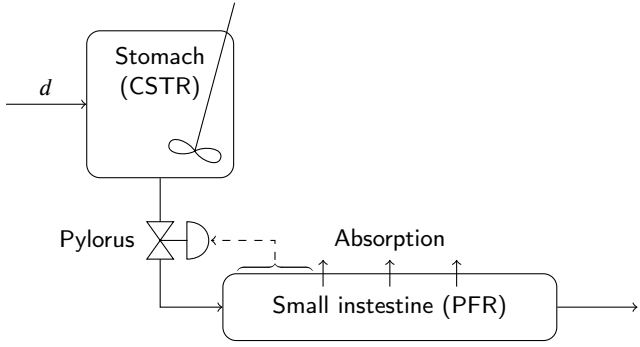
\begin{figure}
    \centering
    \begin{tikzpicture}
    \node[bigbox] (stomach) {};
    \node[text width=40pt, align=center, anchor=north] at ($(stomach.north) - (0pt, 3pt)$) {Stomach (CSTR)};
    
    \node[below=20pt of stomach] (pylorus) {};
    \node[left = 3pt of pylorus] {Pylorus};
    
    \node[longbox, below right=9pt and 20pt of pylorus] (small_intestine) {Small instestine (PFR)};
    
    \draw[->] ($(stomach.west) - (30pt, 0pt)$)              -- (stomach)                                    node [pos=0.5, anchor=south] {$d$};
    \draw[- ] (stomach)                                     -- ($(pylorus.center) + (0pt, \valveheight)$);
    \draw[->] ($(pylorus.center) - (0pt, \valveheight)$)    |- (small_intestine);
    \draw[->] (small_intestine)                             -- ($(small_intestine.east) + (30pt, 0pt)$)     node [pos=0.5, anchor=south] {};
    
    \draw[-, rotate around={-15:(0,0)}] ($(stomach.north east) + (-2.5pt, 10pt)$) -- ++(0pt, -57.5pt)   
    to[out= 45, in=90] ++( 10pt, 0pt) to[out=-90, in=- 45] ++(-10pt, 0pt)                               
    to[out=135, in=90] ++(-10pt, 0pt) to[out=-90, in=-135] ++( 10pt, 0pt);                              
    
    \draw (pylorus.center) -- ++(0.5*\valvewidth, \valveheight) -- ++(-\valvewidth, 0pt) -- (pylorus.center)                            
    -- ++(0.5*\valvewidth, -\valveheight) -- ++(-\valvewidth, 0pt) -- (pylorus.center)                                                  
    -- ++(\valvewidth, 0pt) -- ++(0pt, \valveheight) to[out=0, in=0, looseness=2] ++(0pt, -2*\valveheight) -- ++(0pt, \valveheight);    
    
    \draw[line width=0.17mm, decorate, decoration = {calligraphic brace}] ($(small_intestine.north) + (-57pt, 1pt)$) --  ($(small_intestine.north) + (-29pt, 1pt)$);
    \draw[->, dashed] ($(small_intestine.north) + (-43pt, 5pt)$) |- ($(pylorus.center) + (1.9*\valvewidth, 0pt)$);
    
    \draw[->] ($(small_intestine.north) + (-25pt, -5pt)$) -- ($(small_intestine.north) + (-25pt, 5pt)$);
    \draw[->] ($(small_intestine.north) + (  0pt, -5pt)$) -- ($(small_intestine.north) + (  0pt, 5pt)$) node[anchor=south] {Absorption};
    \draw[->] ($(small_intestine.north) + ( 25pt, -5pt)$) -- ($(small_intestine.north) + ( 25pt, 5pt)$);
\end{tikzpicture}
    \caption{CSTR-PFR model with the feedback mechanism proposed by~\citet{Alskar:etal:2016}.}
    \label{fig:cstr:pfr:diagram}
\end{figure}

The CSTR-PFR model presented here consists of a CSTR representing the stomach and a PFR representing the small intestine, as shown in Fig.~\ref{fig:cstr:pfr:diagram} (see also Section~\ref{sec:models:components:cstr} and~\ref{sec:models:components:pfr}). It is based on the second model presented by~\citet{Moxon:etal:2016}, and we describe three ways of modeling the opening and closing of the pylorus sphincter, which connects the stomach to the small intestine.

The amount of glucose in the stomach, $m_s$, is given by
\begin{subequations}\label{eq:cstr:pfr:stomach}
\begin{align}
    \label{eq:cstr:pfr:stomach:ms}
	\dot m_s &= F_m - F_{sd}, \\
	\label{eq:cstr:pfr:stomach:fsd}
	F_{sd} &= k_{sd} m_s,
\end{align}
\end{subequations}
where $F_m = d$ is the meal input, $F_{sd}$ is the glucose flow rate from the stomach to the duodenum, and $k_{sd}$ is the inverse of a time constant. We either consider $k_{sd}$ to be 1)~constant (the pylorus sphincter is always completely open), 2)~a function of the glucose rate of appearance in the blood (see Section~\ref{sec:cstr:pfr:meal:model:moxon}), or 3)~a function of the amount of glucose in the duodenum (see Section~\ref{sec:cstr:pfr:meal:model:alskar}).
The glucose concentration in the small intestine is described by the PDE
\begin{align}
	\partial_t c_{si} &= -\partial_z N_p - Q_a, & z &\in [z_0, z_f],
\end{align}
where $z$ is the spatial coordinate along the small intestine, and the positions $z_0$ and $z_f$ denote the beginning and end of the small intestine.
The flux $N_p$ describes the peristaltic movement in the small intestine, and it consists of an advection term, $N_{ap}$, and a diffusion term, $N_{dp}$:
\begin{subequations}
\begin{align}
	N_p     &= N_{ap} + N_{dp}, \\
	N_{ap}  &= v_p c_{si}, \\
	N_{dp}  &= -D_p \partial_z c_{si}.
\end{align}
\end{subequations}
The velocity, $v_p$, and the diffusion coefficient, $D_p$, are constant. The glucose absorption, $Q_a$, is given by
\begin{subequations}
\begin{align}
	Q_a &= \frac{2f}{r_{si}} q_a, \\
	q_a &= v_a c_{si},
\end{align}
\end{subequations}
where $r_{si}$ is the radius of the small intestine, and $f$ is a factor describing 1)~the increase in surface area (compared to that of a cylinder) due to villi, microvilli, and plicae circulares, and 2)~the fact that glucose is only absorbed from a fraction of the surface. Furthermore, $v_a$ is the glucose absorption rate.
The flow rate from the stomach to the duodenum is represented as a boundary condition, i.e., the flux at the beginning of the small intestine times the cross-sectional area, $A_{si}$, must equal the glucose flow rate $F_{sd}$:
\begin{align}\label{eq:cstr:pfr:bc}
	A_{si} N_p \rvert_{z = z_0} &= F_{sd}.
\end{align}
Finally, the glucose rate of appearance is the cross-sectional area times the integral of the glucose absorption rate over the length of the small intestine:
\begin{align}
    R_A &= A_{si} \int_{z_0}^{z_f} Q_a \incr z.
\end{align}

\subsubsection{Moxon's feedback mechanism}\label{sec:cstr:pfr:meal:model:moxon}
\citet{Moxon:etal:2017} propose that the glucose flow rate between the stomach and duodenum is equal to 1)~zero if the glucose rate of appearance, $R_A$, is above a certain threshold, $R_{A, max}$, and 2)~$k_{sd}^{max}$ otherwise. This is approximated by
\begin{align}\label{eq:moxon:pyloric:sphincter}
    k_{sd} &= k_{sd}^{max} \frac{1}{1 + \exp(\sigma (R_A - R_{A, max}))},
\end{align}
where the parameter $\sigma$ determines the accuracy of the approximation, i.e., the steepness of $k_{sd}$ around $R_{A, max}$.

\subsubsection{Alsk{\"{a}}r's feedback mechanism}\label{sec:cstr:pfr:meal:model:alskar}
\citet{Alskar:etal:2016} propose that the glucose flow rate can be described using a Hill expression with a high Hill coefficient, $\gamma$, i.e., it approximates an on/off mechanism where the glucose flow rate is equal to zero if the amount of glucose in the duodenum, $m_d$, is above a threshold value, $m_{d, 50}$, and $k_{sd}^{max}$ otherwise. In addition to the original model of the feedback mechanism, we introduce a minimum value, $k_{sd}^{min}$:
\begin{align}\label{eq:alskar:pyloric:sphincter}
	k_{sd} &= k_{sd}^{min} + (k_{sd}^{max} - k_{sd}^{min})\left(1 - \frac{m_d^\gamma}{m_{d, 50}^\gamma + m_d^\gamma}\right).
\end{align}
Finally, the duodenum constitutes the first part of the small intestine (from $z_0$ to $z_d$). Consequently, the amount of glucose in the duodenum is given by
\begin{align}\label{eq:duodenum:glucose}
	m_d &= A_{si} \int_{z_0}^{z_d} c_{si} \incr z.
\end{align}
\begin{remark}
    If $k_{sd}^{min} = 0$ in~\eqref{eq:alskar:pyloric:sphincter}, the glucose flow rate may become close to zero even though the duodenum is almost entirely empty. The reasons are that the velocity of the peristaltic movement, $v_p$, is relatively low and that it is independent of the glucose concentration. Consequently, a very short plug of chyme with a high glucose concentration will move through the duodenum, and once it enters into the jejunum, the duodenum again becomes empty, and the process repeats itself.
\end{remark} 
    \section{Discussion}\label{sec:discussion}
Table~\ref{tab:model:comparison} shows the main characteristics of the models described in Section~\ref{sec:meal:models}: 1)~the types of equations in the model, 2)~the number of states, 3)~whether it is a linear state space model or not, and 4)~whether or not the glucose rate of appearance is linear in the total meal carbohydrate content, $D$. It is more straightforward to simulate models that only contain ODEs. The reason is that PDEs are typically approximated by a set of ODEs using spatial discretization (this is called the method of lines). However, this approximation is derived analytically, and it is problem-specific. In contrast, there exists general-purpose software for simulating models that only contain ODEs. In Appendix~\ref{sec:finite:volume} and~\ref{sec:spectral:galerkin}, we describe two spatial discretization schemes that are relevant to meal models containing PDEs. The approximation often results in a large number of ODEs. Consequently, it is more computationally intensive to simulate models that contain PDEs because the computation time depends strongly on the number of states. Next, linear state space models are simpler to analyze than nonlinear models, and, in important special cases, explicit expressions for their solutions can be derived. Similarly, it can be exploited in both analysis and simulation if a model is linear in the total meal carbohydrate content, $D$. Specifically, if $R_A^{(1)}$ is the glucose rate of appearance over time for $D = 1$, the rate of appearance for any meal carbohydrate content is $R_A^{(1)} D$ if the model is linear in $D$.

Only the CSTR-PFR model contains a PDE (describing the glucose transport in the small intestine). Consequently, when discretized, it will contain more states than the other models, and it will be more computationally intensive to simulate. All the other models contain a small number of states. Furthermore, Hovorka's model and the SIMO model are linear. The CSTR-PFR model is also linear if the pylorus sphincter is modeled as always being open, i.e., if there is no feedback mechanism. The remaining models are nonlinear. Finally, Hovorka's and Dalla Man's models, the SIMO model, and the CSTR-PFR model without feedback are linear in the total meal carbohydrate content, $D$, (see Appendix~\ref{sec:LinearGlucoseRateOfAppearance}).

\begin{figure}
    \centering
    \includegraphics[width=\linewidth]{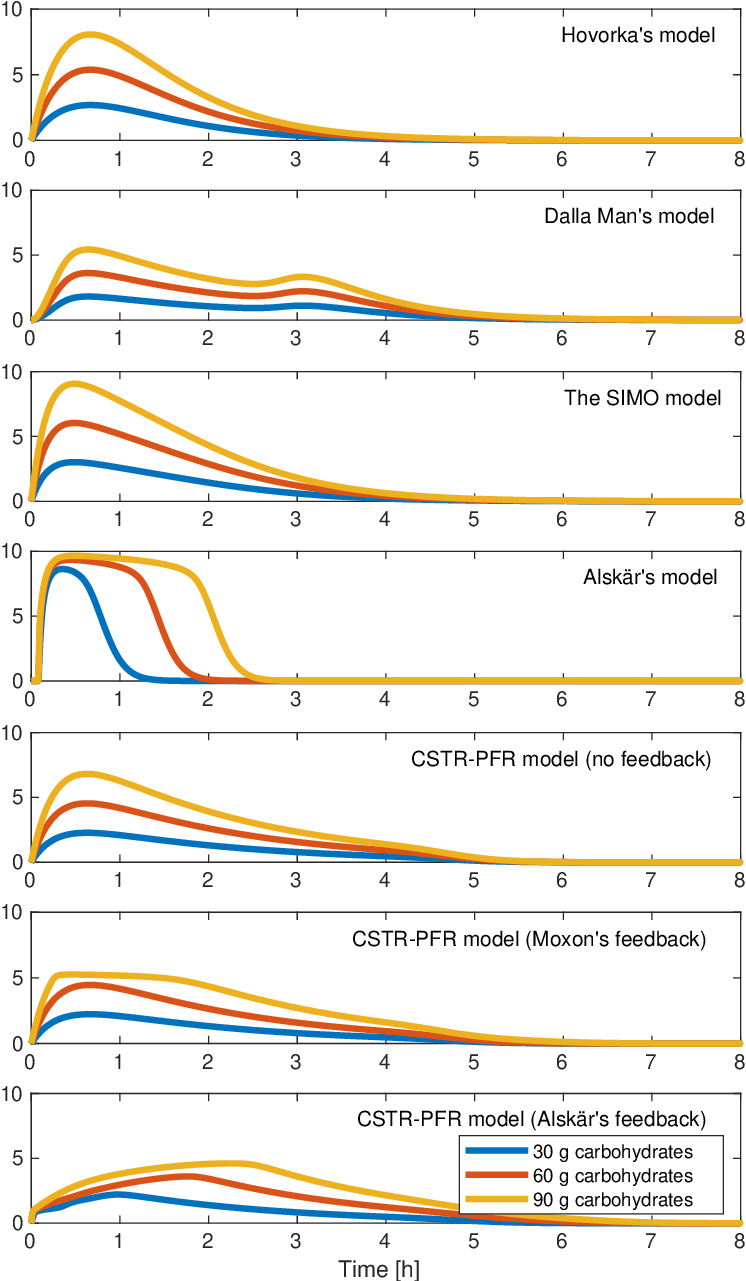}
    \caption{Glucose rate of appearance per body weight for different meal carbohydrate contents.}
    \label{fig:glucoseRateOfAppearanceComparisonMealSizes}
\end{figure}

Fig.~\ref{fig:glucoseRateOfAppearanceComparisonMealSizes} shows the response to meals with different carbohydrate contents. The meal consumption is modeled as instantaneous (as described in Section~\ref{sec:simulation:meal:inputs}). The parameter values used in the various models (see Appendix~\ref{sec:parameter:values}) do not represent the same individual. Therefore, we show the glucose rate of appearance in the blood normalized with body weight. The meal responses predicted by the linear models, i.e., Hovorka's model, the SIMO model, and the CSTR-PFR model without feedback, are qualitatively similar. After an initial rise, the glucose rate of appearance slowly decays to zero. In contrast, Dalla Man's model predicts two peaks. After the initial rise, the rate of appearance decreases and then increases again before decaying to zero. The second peak represents the delayed carbohydrate absorption caused by, e.g., fat and protein in the meal. None of the other models predict more than one peak. In Alsk{\"{a}}r's model, there is a pronounced saturation effect, and the larger meals do not lead to significantly higher glucose absorption rates. Instead, the absorption is prolonged for larger meals. For the largest meal, the CSTR-PFR model using Moxon's feedback mechanism shows a similar saturation effect. However, for the two smaller meals, the saturation threshold is not reached and the simulations are almost identical to those obtained without a feedback mechanism. When Alsk{\"{a}}r's feedback mechanism is used in the CSTR-PFR model, the glucose rate of appearance does not saturate. Instead, after a fast but short rise where the duodenum is filled, it increases slowly. For larger meals, it increases for a longer time. Finally, the simulations clearly demonstrate that the glucose rate of appearance is linear in $D$ for Hovorka's and Dalla Man's models, the SIMO model, and the CSTR-PFR model without feedback.

\begin{figure}
    \centering
    \includegraphics[width=\linewidth]{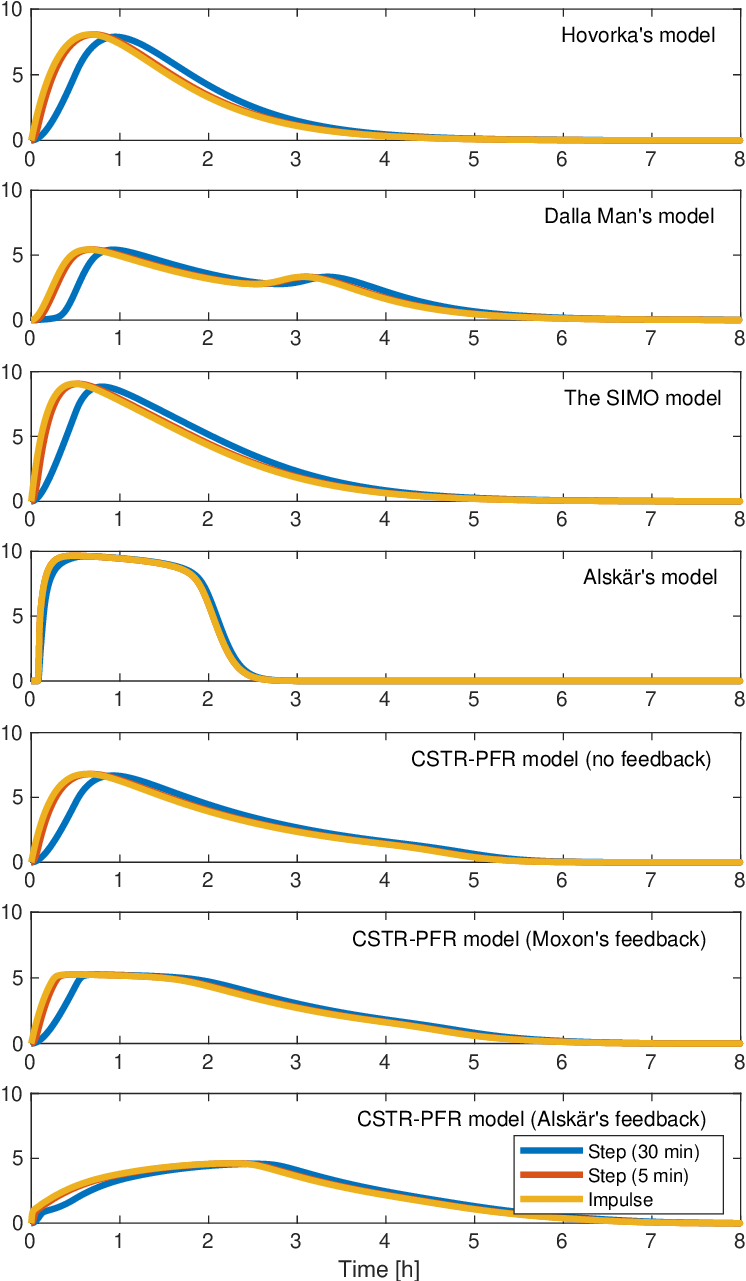}
    \caption{Responses to meals that are consumed instantaneously (i.e., modeled as an impulse function), over 5~min, and over 30~min (i.e., modeled as step functions). The meals contain 90~g carbohydrates.}
    \label{fig:glucoseRateOfAppearanceComparisonImpulseAndStep}
\end{figure}

In Fig.~\ref{fig:glucoseRateOfAppearanceComparisonImpulseAndStep}, we compare the two meal input models discussed in Section~\ref{sec:simulation:meal:inputs}, i.e., 1)~the instantaneous model using an impulse function and 2)~the constant flow rate model using a step function. The two representations are almost identical if the meal is consumed over 5~min. However, there is a pronounced lag for almost all of the models if the meal is consumed over 30~min. The exceptions are Alsk{\"{a}}r's model and the CSTR-PFR model using Alsk{\"{a}}r's feedback mechanism. The reason is that the feedback limits the amount of glucose that can enter into the duodenum. Consequently, the rate at which the stomach is filled has a smaller impact than in the other models.


\begin{table*}
    \centering
    \caption{Main characteristics of the meal models described in Section~\ref{sec:meal:models}. For the CSTR-PFR model, the number of states depends on the discretization of the PDE (resulting in $M$ ODEs). The models are considered linear if they are in the form of a linear state space model~\eqref{eq:linear:state:space:meal:model}. The CSTR-PFR model is linear if $k_{sd}$ in~\eqref{eq:cstr:pfr:stomach:fsd} is constant and nonlinear if it is described by~\eqref{eq:moxon:pyloric:sphincter} or~\eqref{eq:alskar:pyloric:sphincter}. In Appendix~\ref{sec:LinearGlucoseRateOfAppearance}, we show that the glucose rate of appearance is linear in the total meal carbohydrate content, $D$, for some of the models.}
    \label{tab:model:comparison}
    \begin{tabular}{lcccc}
        \hline
         Model & Types of equations & Number of states & Linear & Linear in $D$ \\
         \hline
         \citet{Hovorka:etal:2004}                      & ODEs          & 2        & Yes       & Yes       \\
         \citet{DallaMan:etal:2006, DallaMan:etal:2007} & ODEs          & 3        & No        & Yes       \\
         \citet{DeGaetano:etal:2013}                    & ODEs          & 4        & Yes       & Yes       \\
         \citet{Alskar:etal:2016}                       & ODEs          & 4        & No        & No        \\
         CSTR-PFR~\citep{Moxon:etal:2016}               & ODEs and PDEs & 1 + $M$  & Yes/no    & Yes/no    \\
        \hline
    \end{tabular}
\end{table*}

%
%
    \section{Conclusions}\label{sec:conclusions}
We present a critical discussion of five commonly used mathematical models of gastrointestinal meal glucose absorption. We compare their predictions of the glucose rate of appearance in the blood plasma, and we provide an overview of key aspects of the models, including linearity and the types of equations in the models. The models are relevant to accurate simulation of the metabolism in healthy, diabetic, and obese people, and they can be used to test and develop treatment and prevention strategies.
Furthermore, we discuss general modeling aspects relevant to systematic modeling of meal glucose absorption. Specifically, we discuss model structures, meal input representations, delay approximations, general formulations of stoichiometry and reaction kinetics, and general CSTR and PFR models which can represent different parts and processes in the human metabolism. 
    
    \appendix
    \setcounter{figure}{0}
    \renewcommand\thefigure{A.\arabic{figure}}
    \addcontentsline{toc}{section}{Appendix}
    \counterwithin{figure}{section}
    \numberwithin{equation}{section}
    \numberwithin{figure}{section}
    \numberwithin{table}{section}
    
    \section{Parameter values}\label{sec:parameter:values}
Table~\ref{tab:parameter:values} shows the parameter values used in the simulations presented in Section~\ref{sec:discussion}. Apart from the following exceptions, the parameter values are available in the papers referenced in the table. We use a value of $\sigma$ in Alsk{\"{a}}r's model from an unpublished source, and we choose the value of $k_{sd}$ in the CSTR-PFR model without feedback based on the interval considered by~\citet{Moxon:etal:2016}. In the CSTR-PFR model, $D_p$ and $k_{sd, min}$ were not present in the original model. Therefore, we have chosen the values based on simulations. We have also chosen the value of $\sigma$ in the CSTR-PFR model with Moxon's feedback mechanism. For models where the body weight, $BW$, is not provided, we use a value of 82~kg. Similarly, when $f$ is not provided, we use a value of 1. 
\begin{table*}
\centering
\caption{Parameter values used in Section~\ref{sec:discussion} to simulate the glucose rate of appearance for each of the models described in Section~\ref{sec:meal:models}.}
\label{tab:parameter:values}
\begin{tabular}{lrll}
    \hline
    Symbol & Value & Unit & Description \\
    \hline
    \multicolumn{4}{c}{Hovorka's model~\citep{Hovorka:etal:2004}} \\
    \hline
    $A_G$       & $0.8$   & $-$   & Carbohydrate bioavailability \\
    $\tau_D$    & $40$    & min   & Time constant \\
    $f$         & $1$     & $-$   & Percentage of absorbed glucose that appears in the blood \\
    $BW$        & $82$    & kg    & Body weight \\
    \hline
    \multicolumn{4}{c}{Dalla Man's model~\citep{DallaMan:etal:2006, DallaMan:etal:2007}} \\
    \hline
    $k_{max}$   & $0.0465$    & 1/min & Maximum inverse of gastric emptying time constant \\
    $k_{min}$   & $0.0076$    & 1/min & Minimum inverse of gastric emptying  time constant \\
    $k_{abs}$   & $0.023$     & 1/min & Inverse of intestinal glucose absorption time constant \\
    $k_{gri}$   & $0.0465$    & 1/min & Inverse of grinding (solid-to-liquid in the stomach) time constant \\
    $b$         & $0.69$      & $-$   & Percentage of total meal glucose content corresponding to left inflection point \\
    $c$         & $0.17$      & $-$   & Percentage of total meal glucose content corresponding to right inflection point \\
    $f$         & $0.90$      & $-$   & Percentage of absorbed glucose that appears in the blood \\
    $BW$        & $91$        & kg    & Body weight \\
    \hline
    \multicolumn{4}{c}{The SIMO model~\citep[Table~4]{Panunzi:etal:2020}} \\
    \hline
    $k_{js}$    & $0.026$     & 1/min & Inverse of stomach-to-jejunum time constant \\
    $k_{rj}$    & $0.033$     & 1/min & Inverse of jejunum-to-delay compartment time constant \\
    $k_{lr}$    & $0.030$     & 1/min & Inverse of delay compartment-to-ileum time constant \\
    $k_{gj}$    & $0.036$     & 1/min & Inverse of time constant of glucose absorption in the jejunum \\
    $k_{gl}$    & $0.027$     & 1/min & Inverse of time constant of glucose absorption in the ileum \\
    $f$         & $1$         & $-$   & Percentage of absorbed glucose that appears in the blood \\
    $BW$        & $82$        & kg    & Body weight \\
    \hline
    \multicolumn{4}{c}{Alsk{\"{a}}r's model~\citep{Alskar:etal:2016}} \\
    \hline
    $k_w$           & $0.14$  & 1/min     & Emptying rate for a noncaloric liquid (e.g., water) \\
    $IG_{D50}$      & $7420$  & mg        & Amount of glucose corresponding to a 50\% reduction of the gastric emptying rate \\
    $\gamma$        & $14$    & $-$       & Hill factor \\
    $L_D$           & $0.08$  & $-$       & Length of duodenum relative to the length of the small intestine \\
    $L_J$           & $0.37$  & $-$       & Length of jejunum relative to the length of the small intestine \\
    $T$             & $240$   & min       & The time it takes the chyme to pass through the small intestine \\
    $\sigma$        & $10$    & 1/min     & Parameter in the lag of the gastric emptying rate \\
    $t_{50}$        & $5$     & min       & Delay of the gastric emptying rate \\
    $K_{mG}$        & $6320$  & mg        & Amount of glucose corresponding to a 50\% reduction of the absorption rate \\
    $R_{D, \max}$   & $580$   & mg/min    & Maximum rate of glucose absorption in the duodenum \\
    $R_{J, \max}$   & $2060$  & mg/min    & Maximum rate of glucose absorption in the jejunum \\
    $R_{I, \max}$   & $1330$  & mg/min    & Maximum rate of glucose absorption in the ileum \\
    $F_P$           & $1$     & $-$       & Fraction of glucose absorption that appears in the blood \\
    $BW$            & $82$    & kg        & Body weight \\
    \hline
    \multicolumn{4}{c}{CSTR-PFR model~\citep{Alskar:etal:2016, Moxon:etal:2016, Moxon:etal:2017}} \\
    \hline
    $z_0$                       & $0$         & m         & Position of left end of small intestine \\
    $z_f$                       & $2.85$      & m         & Position of right end of small intestine \\
    $v_p$                       & $0.0102$    & m/min     & Advection velocity of the chyme due to peristaltic movement \\
    $D_p$                       & $0.0001$    & m$^2$/min & Coefficient of glucose diffusion in the chyme \\
    $r_{si}$                    & $0.018$     & m         & Radius of small intestine \\
    $f$                         & $12$        & $-$       & Factor accounting for villi, microvilli, plicae circulares, and effective absorption area \\
    $v_a$                       & $6.4392\cdot10^{-6}$ & m/min & Glucose absorption mass transfer coefficient \\
    $k_{sd}$                    & $0.06$      & 1/min     & Inverse of stomach-to-duodenum time constant (no feedback mechanism) \\
    $k_{sd, max}$               & $0.0554$    & 1/min     & Maximum inverse of stomach-to-duodenum time constant (Moxon's feedback) \\
    $k_{sd, max}$               & $0.14$      & 1/min     & Maximum inverse of stomach-to-duodenum time constant  (Alsk{\"{a}}r's feedback) \\
    $k_{sd, min}$               & $0.0116$    & 1/min     & Minimum inverse of stomach-to-duodenum time constant \\
    $R_{A, max}$                & $420$       & mg/min    & Maximum glucose rate of appearance \\
    $\sigma$                    & $0.1$       & $-$       & Parameter in approximation of Moxon's feedback mechanism \\
    $m_{d, 50}$                 & $7420$      & mg        & Amount of glucose corresponding to a 50\% reduction of the gastric emptying rate \\
    $\gamma$                    & $14$        & $-$       & Hill factor \\
    $BW$                        & $82$        & kg        & Body weight \\
    \hline
\end{tabular}
\end{table*}
    \section{Linearity of the glucose rate of appearance}\label{sec:LinearGlucoseRateOfAppearance}
Here, we show that the glucose rate of appearance in the blood, as a function of time, is linear in the total meal carbohydrate content, $D$, for linear state space models and the model developed by~\citet{DallaMan:etal:2006, DallaMan:etal:2007}, provided that the meal input, $d$, is linear in $D$. That is the case when $d$ is an impulse or step function as described in Section~\ref{sec:simulation:meal:inputs}. Consequently, the meal response can be computed for $D = 1$ and scaled in order to obtain the response for any other value of $D$. For brevity of notation, we do not explicitly indicate the time-dependence of the states.

\begin{remark}
There exist meal input functions, $d$, that are not linear in $D$. For instance, if $d$ is a step function, the meal duration may increase linearly in $D$ while the glucose flow rate remains constant.
\end{remark}

\subsection{Linear models}
The linear meal models are in the form
\begin{subequations}
\begin{align}
    \dot x &= A_c x + B_c d, \\
    y &= C_c x.
\end{align}
\end{subequations}
We introduce the normalized states, inputs, and outputs
\begin{subequations}
\begin{align}
    \tilde x &= x/D, &
    \tilde d &= d/D, &
    \tilde y &= y/D.
\end{align}
\end{subequations}
Note that, by assumption, $\tilde d$ is independent of $D$. Then,
\begin{align}
    \dot{\tilde x} &= \dot x/D = A_c x/D + B_c d/D = A_c \tilde x + B_c \tilde d, \\
    \tilde y &= y/D = C_c x/D = C_c \tilde x,
\end{align}
and, given a simulation of this normalized system, the glucose rate of appearance can be obtained for any meal by scaling the normalized response, i.e., $y = \tilde y D$.

\subsection{Dalla Man model}
As for the linear state space models, we introduce the normalized state variables
\begin{subequations}
\begin{align}
    q_{sto, 1} &= Q_{sto, 1}/D, &
    q_{sto, 2} &= Q_{sto, 2}/D, \\
    q_{gut}    &= Q_{gut}   /D, &
    q_{sto}    &= Q_{sto}   /D,
\end{align}
\end{subequations}
and the normalized meal input $\tilde d = d/D$. Note that $q_{sto} = q_{sto, 1} + q_{sto, 2}$.
The flow rates $R_{12}$ and $R_{gut, pla}$ and the glucose rate of appearance, $R_A$, are linear in their arguments, and they do not depend directly on $D$, i.e.,
\begin{align}
    R_{12}(Q_{sto, 1}) &= R_{12}(q_{sto, 1}) D, \\
    R_{gut, pla}(Q_{gut}) &= R_{gut, pla}(q_{gut}) D, \\
    R_A(Q_{gut}) &= R_A(q_{gut}) D.
\end{align}
In contrast, $R_{sto, gut}$ depends on the gastric emptying rate, $k_{empt} = k_{empt}(Q_{sto}, D)$, which 1)~depends directly on $D$ and 2)~is nonlinear in both its arguments. However, the arguments are not independent, and we show that $k_{empt}$ is independent of $D$ when $Q_{sto} = q_{sto} D$, i.e., that $k_{empt}(q_{sto} D, D) = k_{empt}(q_{sto})$. First, we note that
\begin{align}
    \alpha D &= \frac{5}{2 (1 - b)}, &
    \beta  D &= \frac{5}{2 c}.
\end{align}
Next, using these expressions and substituting $Q_{sto} = q_{sto} D$,
\begin{subequations}
\begin{align}
    \alpha (Q_{sto} - b D)
    &= \alpha (q_{sto} D - b D) = \alpha D (q_{sto} - b) \nonumber \\
    &= \frac{5}{2}\frac{q_{sto} - b}{1 - b}, \\
    \beta (Q_{sto} - c D)
    &= \beta (q_{sto} D - c D) = \beta D (q_{sto} - c) \nonumber \\
    &= \frac{5}{2}\frac{q_{sto} - c}{c}.
\end{align}
\end{subequations}
Finally, we insert into the expression for the gastric emptying rate:
\begin{align}
    k_{empt}
    =&\, k_{min} + \frac{k_{max} - k_{min}}{2}\Bigg(\tanh\left(\alpha (Q_{sto} - b D)\right) \nonumber \\
    &- \tanh\left(\beta (Q_{sto} - c D)\right) + 2\Bigg) \nonumber \\
    =&\, k_{min} + \frac{k_{max} - k_{min}}{2}\Bigg(\tanh\left(\frac{5}{2}\frac{q_{sto} - b}{1 - b}\right) \nonumber \\
    &- \tanh\left(\frac{5}{2}\frac{q_{sto} - c}{c}\right) + 2\Bigg).
\end{align}
Clearly, $k_{empt}$ is independent of $D$.
Consequently, $R_{sto, gut}$ is linear in $D$ for $Q_{sto, 1} = q_{sto, 1} D$ and $Q_{sto, 2} = q_{sto, 2} D$:
\begin{align}
    R_{sto, gut}&(Q_{sto, 1}, Q_{sto, 2}, D) \nonumber \\
    &= k_{empt}(Q_{sto}, D) Q_{sto, 2} \nonumber \\
    &= k_{empt}(q_{sto} D, D) q_{sto, 2} D \nonumber \\
    &= k_{empt}(q_{sto}) q_{sto, 2} D \nonumber \\
    &= R_{sto, gut}(q_{sto, 1}, q_{sto, 2}) D.
\end{align}
In conclusion, the normalized variables are described by
\begin{subequations}
\begin{align}
    \dot q_{sto, 1}
    &=  \dot Q_{sto, 1}/D = d/D - R_{12}(Q_{sto, 1})/D \nonumber \\
    &= \tilde d - R_{12}(q_{sto, 1}), \\
    \dot q_{sto, 2}
    &= \dot Q_{sto, 2}/D \nonumber \\
    &= R_{12}(Q_{sto, 1})/D - R_{sto, gut}(Q_{sto, 1}, Q_{sto, 2}, D)/D \nonumber \\
    &= R_{12}(q_{sto, 1}) - R_{sto, gut}(q_{sto, 1}, q_{sto, 2}), \\
    \dot q_{gut}
    &= \dot Q_{gut}/D \nonumber \\
    &= R_{sto, gut}(Q_{sto, 1}, Q_{sto, 2}, D)/D \nonumber \\
    & \phantom{=} - R_{gut, pla}(Q_{gut})/D \nonumber \\
    &= R_{sto, gut}(q_{sto, 1}, q_{sto, 2}) - R_{gut, pla}(q_{gut}).
\end{align}
\end{subequations}
Given a simulation of this system, the glucose rate of appearance for any meal carbohydrate content, $D$, can be computed as $R_A(q_{gut}) D$.










    \section{Finite volume discretization}\label{sec:finite:volume}
\begin{figure}
	\begin{tikzpicture}
	\foreach \i in {0, ..., 2} {
		\draw (  \i, -0.25) -- (  \i, 0.25); 
		\draw (7-\i, -0.25) -- (7-\i, 0.25); 

		\draw (  \i, -0.25) -- (  \i+1, -0.25); 
		\draw (  \i,  0.25) -- (  \i+1,  0.25); 
		\draw (7-\i, -0.25) -- (7-\i-1, -0.25); 
		\draw (7-\i,  0.25) -- (7-\i-1,  0.25); 

		\node at (  \i, -0.45) {$z_\i$};

		\node at (  \i, -0.8) {$N_\i$};
	}

	\foreach \i in {0, ..., 1} {
		\node at (  \i+0.5, 0.45) {$c_\i$};
	}

	\node at (3.5, 0) {$\cdots$};

	\node at (5.5, 0.45) {$c_{M-2}$};
	\node at (6.5, 0.45) {$c_{M-1}$};

	\node at (5, -0.45) {$z_{M-2}$};
	\node at (6, -0.45) {$z_{M-1}$};
	\node at (7, -0.45) {$z_{M}$};

	\node at (5, -0.8) {$N_{M-2}$};
	\node at (6, -0.8) {$N_{M-1}$};
	\node at (7, -0.8) {$N_{M}$};
\end{tikzpicture}
	\caption{Sketch of the spatially discrete grid used in the finite volume discretization of~\eqref{eq:finite:volume:pde}.}
	\label{fig:grid}
\end{figure}
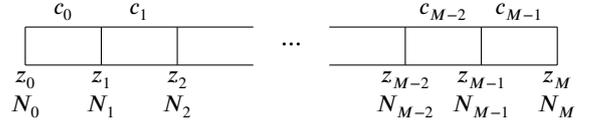

In this appendix, we present a finite volume discretization of the PFR model~\eqref{eq:pfr},
\begin{align}\label{eq:finite:volume:pde}
    \partial_t c &= -\partial_z N + Q,
\end{align}
with the boundary condition
\begin{align}\label{eq:finite:volume:boundary:condition}
    A N \rvert_{z = z_0} &= F.
\end{align}
For simplicity, we ignore the reaction term, $R$, which is treated in the same way as the source term, $Q$.
We discretize the cylindrical domain, $\Omega$, as shown in Fig.~\ref{fig:grid}. That is, we split it into $M$ smaller non-overlapping volumes (also cylinders) such that $\Omega = \bigcup_{i = 0}^{M-1} \Omega_i$, where $\Omega_i = \{s = (z, r, \theta) | z\in[z_i, z_{i+1}], r \in [0, r_c], \theta \in [0, 2\pi[\}$ and $r_c$ is the radius of the cylinder.
First, we integrate over each volume, i.e.,
\begin{align}
	\int_{\Omega_i} \partial_t c \incr s &= -\int_{\Omega_i} \partial_z N \incr s + \int_{\Omega_i} Q \incr s, \nonumber \\
	&= -A \int_{z_i}^{z_{i+1}} \partial_z N \incr z + A \int_{z_i}^{z_{i+1}} Q \incr z,
\end{align}
for $i = 0, \ldots, M-1$. We have interchanged integration and differentiation on the left-hand side, and exploited that the concentration is identical in the plane perpendicular to the motion through the cylinder.
Next, we 1)~use the definition of concentration to define the glucose mass $m_i = \int_{\Omega_i} c \incr s$, 2)~apply Gauss' divergence theorem to the first term on the right-hand side, and 3)~approximate $Q$ as constant in each volume:
\begin{align}
	\dot m_i &= -A (N_{i+1} - N_i) + F_i, & i &= 0, \ldots, M-1.
\end{align}
Here, $N_i$ is an approximation of the flux on the left boundary of the $i$'th volume, and
\begin{align}
	F_i &= A \Delta z_i Q_i, & i &= 0, \ldots, M-1,
\end{align}
where $\Delta z_i = z_{i+1} - z_i$ and $Q_i = Q(c_i)$. Next, we 1)~use the boundary condition~\eqref{eq:finite:volume:boundary:condition}, 2)~use an upwind scheme to approximate the advection term, 3)~use a first-order finite difference approximation of the spatial derivative in the diffusion term, and 4)~assume that there's no diffusion at the end of the cylinder:
\begin{subequations}
\begin{align}
	N_0 &= F/A, \\
	N_i &= N_{a, i} + N_{d, i}, & i &= 1, \ldots, M, \\
	N_{a, i} &= v c_{i-1}, & i &= 1, \ldots, M, \\
	N_{d, i} &= -D_c \frac{c_i - c_{i-1}}{\Delta z_{c, i-1}}, & i &= 1, \ldots, M-1, \\
	N_{d, M} &= 0.
\end{align}
\end{subequations}
The center of the $i$'th volume is
\begin{align}
	z_{c, i} &= z_i + \frac{1}{2} \Delta z_i = z_i + \frac{1}{2}(z_{i+1} - z_i) \nonumber \\
	&= \frac{z_{i+1} + z_i}{2}, \qquad i = 0, \ldots, M-1,
\end{align}
and the distance between the $i$'th and $i+1$'th cell center is
\begin{align}
	\Delta z_{c, i} &= z_{c, i+1} - z_{c, i}, & i &= 0, \ldots, M-2.
\end{align}

For completeness, we also describe the discretization of
\begin{align}
    m_d &= A \int_{z_0}^{z_d} c \incr z,
\end{align}
which is used in the CSTR-PFR model with Alsk{\"{a}}r's feedback mechanism~\eqref{eq:duodenum:glucose}.
Let $K$ be the number of volumes for which $z_{K} \leq z_d < z_{K + 1}$. Then, assuming that the glucose is evenly distributed in each volume,
\begin{align}
	m_d
	&= A\left(\int_{z_K}^{z_d} c \incr z + \sum_{i=0}^{K-1} \int_{z_i}^{z_{i+1}} c \incr z\right) \nonumber \\
	&\approx \frac{z_d - z_{K}}{z_{K + 1} - z_{K}} m_K + \sum_{i=0}^{K-1} m_i.
\end{align}
First, we have split up the integral, and then, we have used the definition of concentration and the assumption of even distribution of the glucose.
    \section{Spectral Galerkin discretization}\label{sec:spectral:galerkin}
In this appendix, we describe a spectral Galerkin discretization~\citep{Kopriva:2009} of the PFR model~\eqref{eq:pfr}. As with the finite volume discretization, we disregard the reaction term, $R$, because it is discretized in the same way as the source term, $Q$. That is, we discretize the system
\begin{subequations}\label{eq:general:pde}
	\begin{align}
		\partial_t c &= -\partial_z N + Q, \\
		A N \rvert_{z = z_0} &= F.
	\end{align}
\end{subequations}
In this appendix, we assume that $z\in[-1, 1]$ (referred to as the computational domain). That is typically not the case. However, the actual physical domain can be mapped onto the computational domain and the system can be transformed accordingly, as described in Appendix~\ref{sec:affine:domain:transformation}.

We approximate the solution, $c$, as a sum of products between time-dependent functions and space-dependent polynomials:
\begin{align}\label{eq:spectral:galerkin:approximate:concentration}
	c(t, z)
	&\approx \hat c(t, z) = \sum_{m=0}^M \hat c_m \ell_m.
\end{align}
Here, $\hat c_m = \hat c_m(t) = \hat c(t, z_m)$ is the $m$'th time-dependent coefficient, $\{z_m\}_{m=0}^M$ is a set of collocation points, and $\ell_m = \ell_m(z)$ is the $m$'th Lagrange polynomial (see Appendix~\ref{sec:spectral:galerkin:lagrange:polynomials}). An important property of such polynomials is that $\ell_m(z_i) = \delta_{im}$, i.e., it is equal to one when evaluated in the $m$'th collocation point and zero when evaluated in any other collocation point.
For brevity of notation, we often omit the dependency on time and space when functions are evaluated at $t$ and $z$.
Furthermore, for clarity, we explicitly indicate the arguments of the flux, $N$, and the source term, $Q$.

We require that the approximate solution, $\hat c$, satisfies the PDE weakly. Specifically,
\begin{align}\label{eq:spectral:galerkin:weak:form}
	\int_{z_0}^{z_f} \left(\partial_t \hat c + \partial_z N(\hat c) - Q(\hat c)\right) \phi \incr z &= 0
\end{align}
must be satisfied for any test function $\phi$ that belongs to the same function space as the approximate solution, i.e., for any polynomial.
As for the solution, we write the test function as a Lagrange polynomial:
\begin{align}\label{eq:spectral:galerkin:weak:form:test:polynomial}
	\phi(t, z) &= \sum_{n=0}^M \phi_n \ell_n.
\end{align}
Here, $\phi_n = \phi_n(t) = \phi_n(t, z_n)$ is the $n$'th time-dependent coefficient. We substitute the expression for the test function in~\eqref{eq:spectral:galerkin:weak:form}:
\begin{align}\label{eq:spectral:galerkin:weak:form:2}
	\sum_{n=0}^M \phi_n \int_{z_0}^{z_f} (\partial_t \hat c + \partial_z N(\hat c) - Q(\hat c)) \ell_n \incr z &= 0.
\end{align}
This equality must be satisfied for any polynomial $\phi$, which means that it must be satisfied for any combination of values of $\{\phi_n\}_{n=0}^m$ at any point in time. Therefore, the integral must equal zero, i.e.,
\begin{align}\label{eq:spectral:galerkin:weak:form:3}
	\int_{z_0}^{z_f} (\partial_t \hat c + \partial_z N(\hat c) - Q(\hat c)) \ell_n \incr z &= 0,
\end{align}
for $n = 0, \ldots, M$. Next, we use integration by parts to rewrite the integral of the flux term. The result is
\begin{align}
	\int_{z_0}^{z_f} \partial_z N(\hat c) \ell_n \incr z =\,& \left[N(\hat c) \ell_n\right]_{z_0}^{z_f} \nonumber \\
	&- \int_{z_0}^{z_f} N(\hat c) \diff{\ell_n}{z} \incr z,
\end{align}
for $n = 0, \ldots, M$. We insert this expression and the expression for the approximate concentration~\eqref{eq:spectral:galerkin:approximate:concentration} into~\eqref{eq:spectral:galerkin:weak:form:3} in order to obtain
\begin{align}\label{eq:spectral:galerkin:before:quadrature}
	&\sum_{m=0}^M \diff{\hat c_m}{t} \int_{z_0}^{z_f} \ell_m \ell_n \incr z + \left[N(\hat c) \ell_n\right]_{z_0}^{z_f} \nonumber \\
	&- \int_{z_0}^{z_f} N(\hat c) \diff{\ell_n}{z} \incr z - \int_{z_0}^{z_f} Q(\hat c) \ell_n \incr z = 0,
\end{align}
for $n = 0, \ldots, M$. We approximate the integrals using quadrature (see Appendix~\ref{sec:spectral:galerkin:jacobi:polynomials}). Consequently,
\begin{align}
	&\sum_{m=0}^M \diff{\hat c_m}{t} \sum_{l=0}^M \ell_m(z_l) \ell_n(z_l) w_l + \left[N(\hat c) \ell_n\right]_{z_0}^{z_f} \nonumber \\
	&- \sum_{l=0}^M N(\hat c(t, z_l)) \diff{\ell_n}{z}(z_l) w_l \nonumber \\
	&- \sum_{l=0}^M Q(\hat c(t, z_l)) \ell_n(z_l) w_l = 0,
\end{align}
for $n = 0, \ldots, M$. We exploit that $\ell_m(z_l) = \delta_{ml}$ and that $\hat c(t, z_l) = \hat c_l(t)$:
\begin{align}
	&\diff{\hat c_n}{t} w_n + \left[N(\hat c) \ell_n\right]_{z_0}^{z_f} - \sum_{l=0}^M N(\hat c_l) \diff{\ell_n}{z}(z_l) w_l \nonumber \\
	&- Q(\hat c_n) w_n = 0, \quad n = 0, \ldots, M.
\end{align}
Finally, we rearrange terms in order to obtain the ODEs for each of the coefficients,
\begin{align}\label{eq:spectral:galerkin:final}
	\diff{\hat c_n}{t} =\,& -\frac{1}{w_n} \left[N(\hat c) \ell_n(z)\right]_{z_0}^{z_f} \nonumber \\
	&+ \frac{1}{w_n} \sum_{l=0}^M N(\hat c_l) \diff{\ell_n}{z}(z_l) w_l + Q(\hat c_n),
\end{align}
for $n = 0, \ldots, M$.
\begin{remark}
When using Gauss-Lobatto quadrature, the boundary contribution in~\eqref{eq:spectral:galerkin:final} (i.e., the first term on the right-hand side) is only nonzero for the boundary coefficient (i.e., for $n=0$) because $z_0$ and $z_f$ are collocation points. For Gauss quadrature, the boundary contribution is nonzero for all of the differential equations because $z_0$ and $z_f$ are not collocation points.
\end{remark}

\subsection{Jacobi polynomials and quadrature}\label{sec:spectral:galerkin:jacobi:polynomials}
We denote by $P_k^{(\alpha, \beta)}$ a general $k$'th order Jacobi polynomial~\citep{Kopriva:2009}, and we describe two special cases: 1)~Legendre polynomials and 2)~Chebyshev polynomials.
Both polynomials can be used in a Gauss or Gauss-Lobatto quadrature rule:
\begin{align}
    \int_{z_0}^{z_f} f(z) w(z) \incr z \approx \sum_{l=0}^M f(z_l) w_l.
\end{align}
The weight function $w$ and the weights $\{w_l\}_{l=0}^M$ are specific to each Jacobi polynomial. Gauss quadrature rules are exact for polynomials of up to order $2 M + 1$, but do not include the endpoints, $z_0$ and $z_f$. The endpoints are included in Gauss-Lobatto quadrature rules, which are only exact for polynomials of up to order $2 M - 1$.
\begin{remark}
    The Sturm-Liouville problem consists of the following differential equation combined with boundary conditions on $u$ (not shown).
    \begin{align}
    	-\diff{}{z}\left(p(z) \diff{u}{z}\right) + q(z) u &= \lambda w(z) u, & a &< z < b.
    \end{align}
    Jacobi polynomials, $P_k^{(\alpha, \beta)}(z)$, are eigenfunctions of the specific Sturm-Liouville problem
    \begin{align}
    	-\diff{}{z}\left((1 - z)^{1 + \alpha} (1 + z)^{1 + \beta} \diff{u}{z}\right) \nonumber \\
    	= \lambda (1 - z)^{\alpha} (1 + z)^{\beta} u,
    \end{align}
    where $\alpha, \beta > -1$ and $-1 < z < 1$.
\end{remark}

\subsubsection{Legendre polynomials}
The $k$'th order Legendre polynomial, $L_k = P_k^{(0, 0)}$, is obtained with $\alpha = \beta = 0$, and it is defined recursively starting with $L_0(z) = 1$ and $L_1(z) = z$. Subsequently,
\begin{align}\label{eq:legendre}
    L_{k+1}(z) &= \frac{2k+1}{k+1} z L_k(z) - \frac{k}{k+1} L_{k-1}(z),
\end{align}
and the weight function is
\begin{align}\label{eq:legendre:w}
	w(z) &= 1.
\end{align}
The Legendre polynomials also satisfy
\begin{align}
    (2k + 1) L_k(z) &= \diff{L_{k+1}}{z}(z) - \diff{L_{k-1}}{z}(z).
\end{align}
The Legendre Gauss collocation points, $\{z_l\}_{l=0}^M$, are the zeros of $L_{M+1}$, and the weights are
\begin{align}
	w_l &= \frac{2}{(1 - z_l^2)\left(\diff{L_{M+1}}{z}(z_l)\right)^2}, & l &= 0, \ldots, M.
\end{align}
The Legendre Gauss-Lobatto collocation points, $\{z_l\}_{l=0}^M$, are -1, 1, and the zeros of $\diff{L_M}{z}$, and the weights are
\begin{align}\label{eq:quadrature:gauss:lobatto:legendre:w}
	w_l &= \frac{2}{M(M + 1)} \frac{1}{\left(L_M(z_l)\right)^2}, & l &= 0, \ldots, M.
\end{align}

\subsubsection{Chebyshev polynomials}
For Chebyshev polynomials, $\alpha = \beta = -1/2$ and the $k$'th order polynomial is denoted by $T_k = P_k^{(-1/2, -1/2)}$. Chebyshev polynomials are given by the explicit expression
\begin{align}
	T_k(z) &= \cos\left(k \cos^{-1}(z)\right),
\end{align}
but they also satisfy a recursion. It starts with $T_0(z) = 1$ and $T_1(z) = z$ and is followed by
\begin{align}
	T_{k+1}(z) &= 2 z T_k(z) - T_{k-1}(z).
\end{align}
They also satisfy
\begin{align}
	2 T_k(z) &= \frac{1}{k+1} \diff{T_{k+1}}{z}(z) - \frac{1}{k-1} \diff{T_{k-1}}{z}(z),
\end{align}
and the weight function is
\begin{align}
	w(z) &= \frac{1}{\sqrt{1 - z^2}}.
\end{align}
The Chebyshev Gauss collocation points and weights are given by
\begin{subequations}
	\begin{align}
		z_l &= \cos\left(\frac{2l + 1}{2M + 2} \pi\right), & l &= 0, \ldots, M, \\
		w_l &= \frac{\pi}{M + 1}, & l &= 0, \ldots, M,
	\end{align}
\end{subequations}
and the Chebyshev Gauss-Lobatto collocation points and weights are
\begin{subequations}\label{eq:quadrature:gauss:lobatto:chebyshev}
	\begin{align}
		\label{eq:quadrature:gauss:lobatto:chebyshev:x}
		z_l &= \cos\left(\frac{l\pi}{M}\right), \\
		\label{eq:quadrature:gauss:lobatto:chebyshev:w}
		w_l &=
		\begin{cases}
			\frac{\pi}{2 M}, & l \in \{0, M\}, \\
			\frac{\pi}{M}, & l = 1, \ldots, M-1,
		\end{cases}
	\end{align}
\end{subequations}
for $l = 0, \ldots, M$.
\subsection{Lagrange polynomials}\label{sec:spectral:galerkin:lagrange:polynomials}
Here, we describe the Lagrange polynomials which we use several times in the derivation of the spectral Galerkin method presented in this section.
For arbitrary $z$, the $m$'th Lagrange polynomial of order $M+1$ and its derivatives are~\citep{Berrut:Trefethen:2004}
\begin{subequations}
	\begin{align}
		\ell_m(z) &= \prod_{\substack{l = 0\\l \neq m}}^M \frac{z - z_l}{z_m - z_l} = \frac{1}{s(z)} \frac{\tilde w_m}{z - z_m}, \\
		\diff{\ell_m}{z}(z) &= \frac{1}{s(z)}\left(\frac{-\tilde w_m}{(z - z_m)^2} - \ell_m(z) \diff{s}{z}(z)\right), \\
		\ndiff[2]{\ell_m}{z}(z)
		&= \frac{1}{s(z)}\bigg(\frac{2 \tilde w_m}{(z - z_m)^3} - 2 \diff{\ell_m}{z}(z) \diff{s}{z}(z) \nonumber \\
		&- \ell_m(z) \ndiff[2]{s}{z}(z)\bigg),
	\end{align}
\end{subequations}
where the corresponding weight is
\begin{align}
	\tilde w_m &= \prod_{\substack{l = 0\\l \neq m}}^M \frac{1}{z_m - z_l},
\end{align}
and the auxiliary function, $s$, and its derivatives are given by
\begin{subequations}
	\begin{align}
		s(z) &= \sum_{l=0}^M \frac{\tilde w_l}{z - z_l}, \\
		\diff{s}{z}(z) &= \sum_{l=0}^M \frac{-\tilde w_l}{(z - z_l)^2}, \\
		\ndiff[2]{s}{z}(z) &= \sum_{l=0}^M \frac{2 \tilde w_l}{(z - z_l)^3}.
	\end{align}
\end{subequations}
The above expressions for the derivative of $\ell_m$ cannot be evaluated in the collocation points, $\{z_m\}_{m=0}^M$, because it would result in division by zero. In the collocation points, the Lagrange polynomials and their derivatives are
\begin{subequations}
	\begin{align}
		\ell_m(z_l) &= \delta_{ml}, \\
		\diff{\ell_m}{z}(z_l) &=
		\begin{cases}
			\frac{\tilde w_m}{\tilde w_l (z_l - z_m)}, & l \neq m, \\
			-\sum\limits_{\substack{j = 0\\j\neq l}}^M \diff{\ell_j}{z}(z_i), & l = m,
		\end{cases} \\
		\ndiff[2]{\ell_m}{z}(z_l) &=
		\begin{cases}
			-2 \diff{\ell_m}{z}(z_l)\left(-\diff{\ell_l}{z}(z_l) + \frac{1}{z_l - z_m}\right), & l \neq m, \\
			-\sum\limits_{\substack{j = 0\\j\neq l}}^M \ndiff[2]{\ell_j}{z}(z_i), & l = m.
		\end{cases}
	\end{align}
\end{subequations}
\begin{remark}
There is a sign error in the last term in the parenthesis on the right-hand side of~(9.4) in the paper by~\cite{Berrut:Trefethen:2004}.
\end{remark}
\subsection{Domain transformation}\label{sec:affine:domain:transformation}
If the physical spatial domain is not $[z_0, z_f] = [-1, 1]$, we transform the system by introducing the spatial coordinate $\xi = \xi(z) = 2 \frac{z - z_0}{z_f - z_0} - 1 \in [-1, 1]$. We use the inverse transformation, $z = z(\xi) = \frac{1}{2}(\xi + 1)(z_f - z_0) + z_0$, to express the PDE and the boundary condition in terms of $\xi$. For simplicity, in this appendix, we assume that the velocity, $v$, and the diffusion coefficient, $D_c$, are independent of space, $z$, and concentration, $c$. First, we use the chain rule to derive the partial derivatives of the concentration with respect to $\xi$:
\begin{subequations}
	\begin{align}
		\partial_\xi c
		&= \partial_z c \diff{z}{\xi}, \\
		\partial_{\xi\xi} c
		&= \partial_{zz} c \left(\diff{z}{\xi}\right)^2 + \partial_z c \ndiff[2]{z}{\xi} = \partial_{zz} c \left(\diff{z}{\xi}\right)^2.
	\end{align}
\end{subequations}
We have exploited that $z$ is linear in $\xi$, i.e., $\ndiff[2]{z}{\xi} = 0$. Consequently,
\begin{align}
	\partial_z c &= \partial_\xi c \left(\diff{z}{\xi}\right)^{-1}, &
	\partial_{zz} c &= \partial_{\xi\xi} c \left(\diff{z}{\xi}\right)^{-2}.
\end{align}
Next, we use the chain rule to express the partial derivatives of the flux:
\begin{subequations}
	\begin{align}
		\partial_\xi N
		&= \partial_z N \diff{z}{\xi}
		= \left(v \partial_z c + \partial_z J\right) \diff{z}{\xi} 
		= v \partial_\xi c + \partial_\xi J, \\
		\partial_\xi J
		&= \partial_z J \diff{z}{\xi}
		= - D_c \partial_{zz} c \diff{z}{\xi} 
		= -D_c \partial_{\xi\xi} c\left(\diff{z}{\xi}\right)^{-1}.
	\end{align}
\end{subequations}
Consequently,
\begin{align}
	\partial_t c
	&= -\partial_z N + Q = -\partial_\xi N \left(\diff{z}{\xi}\right)^{-1} + Q,
\end{align}
and
\begin{subequations}
	\begin{align}
		\partial_\xi N \left(\diff{z}{\xi}\right)^{-1} &= v \left(\diff{z}{\xi}\right)^{-1} \partial_\xi c + \partial_\xi J \left(\diff{z}{\xi}\right)^{-1}, \\
		\partial_\xi J \left(\diff{z}{\xi}\right)^{-1}
		&= -D_c \left(\diff{z}{\xi}\right)^{-2} \partial_{\xi\xi} c.
	\end{align}
\end{subequations}
Therefore, the transformed system
\begin{align}
    \partial_t c &= -\partial_\xi \bar N + Q,
\end{align}
where
\begin{subequations}
	\begin{align}
		\bar N &= \bar v c + \bar J, \\
		\bar J &= -\bar D_c \partial_\xi c, \\
		\bar v &= v \left(\diff{z}{\xi}\right)^{-1}, \\
		\bar D_c &= D_c \left(\diff{z}{\xi}\right)^{-2},
	\end{align}
\end{subequations}
is in the form~\eqref{eq:general:pde}, and $\xi \in [-1, 1]$. The two main differences between the original and the transformed system are the velocity and the diffusion coefficient. Note that we have exploited that $\diff{z}{\xi}$ is independent of $\xi$ (i.e., it is constant).
The boundary condition in the transformed system is
\begin{align}\label{eq:spectral:galerkin:transformed:boundary:condition}
	A \tilde N \rvert_{z = z_0} &= F,
\end{align}
where
\begin{subequations}
	\begin{align}
		\tilde N &= v c + \tilde J, \\
		\tilde J &= -D_c \partial_z c = -D_c \partial_\xi c \left(\diff{z}{\xi}\right)^{-1} = -\tilde D_c \partial_\xi c, \\
		\tilde D_c &= D_c \left(\diff{z}{\xi}\right)^{-1}.
	\end{align}
\end{subequations}
Finally, we reformulate the integral in~\eqref{eq:duodenum:glucose} in the CSTR-PFR model with Alsk{\"{a}}r's feedback mechanism. We use that $\incr z = \diff{z}{\xi} \incr \xi$:
\begin{align}
	m_d &= A \int_{z_0}^{z_d} c \incr z = A \int_{\xi_0}^{\xi_d} c \diff{z}{\xi} \incr \xi,
\end{align}
where $\xi_0 = -1$ and $\xi_d = \xi(z_d)$.
\begin{remark}
    As $\diff{z}{\xi}$ is constant, the boundary condition~\eqref{eq:spectral:galerkin:transformed:boundary:condition} can also be formulated in terms of the transformed flux $\bar N$, i.e.,
    \begin{align}
        A \bar N \rvert_{z=z_0} = \bar F,
    \end{align}
    where
    \begin{align}
        \bar F = F \left(\diff{z}{\xi}\right)^{-1}.
    \end{align}
\end{remark} 
    
\bibliographystyle{cas-model2-names}

\bibliography{./references/mealmodels}
\end{document}